\documentclass{emulateapj}
\usepackage{times}
\usepackage{hyperref}
\newcommand{\beq}{\begin{equation}}
\newcommand{\eeq}{\end{equation}}
\def\alp{\mbox{$\alpha$}}

\def\farcs{\hbox{$.\!\!^{\prime\prime}$}}
\def\farchs{\hbox{$.\!\!^{\rm s}$}}

\def\arcmin{\hbox{$^\prime$}}

\def\solar{\mbox{$_{\normalsize\odot}$}}

\def\deg{\hbox{$^\circ$}}
\newcommand{\AmS}{{\protect\the\textfont2
  A\kern-.1667em\lower.5ex\hbox{M}\kern-.125emS}}
\newcommand{\lsim}{\ \raise
-2.truept\hbox{\rlap{\hbox{$\sim$}}\raise5.truept\hbox{$<$}\ }}
\newcommand{\gsim}{\ \raise
-2.truept\hbox{\rlap{\hbox{$\sim$}}\raise5.truept\hbox{$>$}\ }}
\newcommand{\simsim}{\ \raise
-2.truept\hbox{\rlap{\hbox{$\sim$}}\raise5.truept\hbox{$\sim$}\ }}

\slugcomment{Accepted for publication in the Astrophysical Journal}

\righthead{Pre--Main-Sequence Stars across Shapley Constellation III in the LMC}
\lefthead{Gouliermis, et al.}

\shorttitle{Pre--Main-Sequence Stars across Shapley Constellation III in the LMC}
\shortauthors{Gouliermis, et al.}

\begin{document}
 
%\title{Variations of Pre-Main Sequence Populations in star-forming regions \\
%across Shapley Constellation III in the Large Magellanic Cloud\altaffilmark{1,}\altaffilmark{2}}

%\title{Variations of Pre-Main Sequence Populations across Shapley Constellation III in the LMC\altaffilmark{1,}\altaffilmark{2}}

\title{Pre--Main-Sequence stellar populations across Shapley Constellation III.\\ 
%in the LMC with {\em HST}~WFPC2 Imaging. 
I. Photometric Analysis and Identification\altaffilmark{1,}\altaffilmark{2}}

%\footnotemark[1]
%\footnotetext[1]{}

%% Use \author, \affil, and the \and command to format
%% author and affiliation information.
%% Note that \email has replaced the old \authoremail command
%% from AASTeX v4.0. You can use \email to mark an email address
%% anywhere in the paper, not just in the front matter.
%% As in the title, you can use \\ to force line breaks.

\author{Dimitrios A. Gouliermis\altaffilmark{3}, 
        Andrew E. Dolphin\altaffilmark{4}, 
        Massimo Robberto\altaffilmark{5}, 
        Robert A. Gruendl\altaffilmark{6},
        You-Hua Chu\altaffilmark{6},
        Mario Gennaro\altaffilmark{3}, 
        Thomas Henning\altaffilmark{3},
        Michael Rosa\altaffilmark{7,}\altaffilmark{8},
        Nicola Da Rio\altaffilmark{3,5}, 
        Wolfgang Brandner\altaffilmark{3}, 
        Martino Romaniello\altaffilmark{7},
        Guido De~Marchi\altaffilmark{9},
        Nino Panagia\altaffilmark{5,10,11}, 
        and 
        Hans Zinnecker\altaffilmark{12}
        }

%\author{D. A. Gouliermis\altaffilmark{3}, 
%        A. E. Dolphin\altaffilmark{4}, 
%        M. Robberto\altaffilmark{5}, 
%        M. R. Rosa\altaffilmark{6,}\altaffilmark{7},
%        W. Brandner\altaffilmark{3}, 
%        Y.-H. Chu\altaffilmark{8},
%        G. De~Marchi\altaffilmark{9},
%        R. A. Gruendl\altaffilmark{8},
%        Th. Henning\altaffilmark{3},
%        N. Panagia\altaffilmark{5,10,11}, 
%        M. Romaniello\altaffilmark{12},
%        and
%        H. Zinnecker\altaffilmark{13}
%}

\altaffiltext{1}{Based on observations made with the NASA/ESA {\em 
Hubble Space Telescope}, obtained at the Space Telescope Science 
Institute, which is operated by the Association of Universities for 
Research in Astronomy, Inc. under NASA contract NAS 5-26555.}

\altaffiltext{2}{Research supported by the National Aeronautics and 
Space Administration (NASA), and the German Aerospace 
Center (DLR).}

\altaffiltext{3}{Max Planck Institute for Astronomy, K\"{o}nigstuhl 
17, 69117 Heidelberg, Germany}

\email{dgoulier@mpia-hd.mpg.de}

\altaffiltext{4}{Raytheon Company, PO Box 11337, Tucson, AZ 85734, USA}

\altaffiltext{5}{Space Telescope Science Institute, 3700 San Martin Dr., Baltimore MD 21218, USA}

\altaffiltext{6}{Department of Astronomy, University of Illinois, 1002
West Green Street, Urbana, IL 61801, USA}

\altaffiltext{7}{ESO, Karl-Schwarzschild-Str. 2, 85748 Garching, Germany}

%\altaffiltext{7}{ST-ECF, ESO, Karl-Schwarzschild-Str. 2, 85748 Garching, Germany}

\altaffiltext{8}{Affiliated with the Space Telescope Operations Division, RSSD, ESA}

\altaffiltext{9}{ESA, Space Science Department, Keplerlaan 1, 2200 AG Noordwijk, the Netherlands}

\altaffiltext{10}{INAF - Osservatorio Astrofisico di Catania, Via Santa Sofia 78, I-95123 Catania, Italy}

\altaffiltext{11}{Supernova Ltd., OYV \#131, Northsound Road, Virgin Gorda, British Virgin Islands}

\altaffiltext{12}{Institut f\"{u}r Raumfahrtsyteme, Universit|\"{a}t Stuttgart, Pfaffenwaldring 31, 70569 Stuttgart, Germany}

%Astrophysikalisches Institut Potsdam, An der Sternwarte 16 14482, Potsdam, Germany
%\email{zinnecker@dsi.uni-stuttgart.de}

\begin{abstract} 

We present our investigation of pre--main-sequence (PMS) stellar populations in the Large Magellanic Cloud (LMC) 
from imaging with Hubble Space Telescope WFPC2 camera. Our targets of interest are four star-forming regions located 
at the periphery of the super-giant shell LMC 4 (Shapley Constellation III). The PMS stellar content of the regions is revealed 
through the differential Hess diagrams and the observed color-magnitude diagrams (CMDs). Further statistical analysis of 
stellar distributions along cross-sections of the faint part of the CMDs allowed the quantitative assessment of the PMS stars 
census, and the isolation of faint PMS stars as the true low-mass stellar members of the regions. These distributions are found 
to be well represented by a double Gaussian function, the first component of which represents the main-sequence field stars 
and the second the native PMS stars of each region. Based on this result, a cluster membership probability was assigned to 
each PMS star according to its CMD position. The higher extinction in the region LH 88 did not allow the unambiguous 
identification of its native  stellar population. The CMD distributions of the PMS stars with the highest membership probability 
in the regions LH 60, LH 63 and LH 72 exhibit an extraordinary similarity among the regions, suggesting that these stars share 
common characteristics, as well as common recent star formation history. Considering that the regions are located at different 
areas of the edge of LMC 4, this finding suggests that star formation along the super-giant shell may have occurred almost 
simultaneously.

\end{abstract}


\section{Introduction}

Considerable star formation in the Large Magellanic Cloud  (LMC) takes place in interstellar shells, 
ranging from small bubbles to large super-giant shells \citep[see, e.g.,][]{chu09}. Among the latter, LMC~4 
\citep{meaburn80}  is the largest super-giant shell in the Local Group, encompassing a remarkable {\sc Hi} 
cavity of diameter $\sim$~1.9~kpc \citep{dopita85}. Centered on {\sl Shapley Constellation III} \citep{shapley51} 
the area of LMC~4 comprises over 500 clusters, associations and emission nebulae \citep{bica99}. 
In the Milky Way, the youngest stellar associations with $\tau < 10$~Myr are surrounded by bright {\sc Hii} regions, and  
comprise high-mass main sequence (MS) stars as well as intermediate- and low-mass pre-main 
sequence (PMS) stars. While the high-mass (OB-type) MS and intermediate-mass (Herbig Ae/Be) PMS 
populations are the direct signature of the youthfulness of their hosting associations, the low-mass PMS 
stars preserve a record of the complete recent star formation history of the region over long periods, since their evolution 
is extremely slow and can last up to many tens of Myr\footnote{Typical contraction time for a 1~M{\solar} 
star is 50~Myr and for a 0.5~M{\solar} star 200~Myr \citep[][]{karttunen07}.}. Bearing this in mind, 
we undertake a research project that aims at a comprehensive study of the stellar populations in LMC 
star-forming regions, with emphasis on the recent star formation history in the vicinity of LMC~4 as 
recorded in the low-mass PMS populations. 

%\clearpage
%%%%%%%%%%%%%%%%%%%%%%%%%%%%%%%%%%%%%%%%%%%%%%%%%%%%%%%%%%%%
\begin{deluxetable*}{cccccccc}
\tabletypesize{\footnotesize}
\label{t:obs}
%\tablecolumns{8}
\tablewidth{0pc}
%\tablenum{1}
\tablecaption{Description of the WFPC2 observations within our HST Program GO-11547. \label{t:obs} }
\tablehead{
\colhead{Association} & 
\colhead{Visit} & 
\colhead{R.A.} &
\colhead{Decl.} &
\colhead{Data set} &
\colhead{Filter} &
\colhead{Exposure} &
\colhead{Observation} \\
\colhead{(Field)} & 
\colhead{No} &
\colhead{(J2000.0)} &
\colhead{(J2000.0)} &
\colhead{filename} &
\colhead{} &
\colhead{time (s)} &
\colhead{Date} 
} 
\startdata
LH~63 & 1 &  05$^{\rm h}$27$^{\rm m}$47\farchs91& $-$67\deg25\arcmin43\farcs0& ub0h010 & F300W & $1 \times ~10$, $6 \times 160$ & 2008 Jul 15\\
             &    &                                                                       &                                                         &                 & F450W & $1 \times ~10$, $6 \times 160$ & 2008 Jul 15\\
             &    &                                                                       &                                                         &                 &  F555W & $4 \times 300$, $2 \times 350$ & 2008 Jul 15\\
             &    &                                                                       &                                                         &                 & F656N & $1 \times 300$, $1 \times 260$, $1 \times 800$ & 2008 Jul 15\\
             &    &                                                                       &                                                         &                 & F814W & $4 \times 260$, $2 \times 300$ & 2008 Jul 15\\
             & 2 &  05$^{\rm h}$28$^{\rm m}$05\farchs60& $-$67\deg25\arcmin35\farcs1& ub0h020 & F555W & $4 \times 300$, $2 \times 350$ & 2008 Jul 19\\                     
             &   &                                                                       &                                                         &                 & F814W & $4 \times 260$, $2 \times 300$ & 2008 Jul 19\\
LH~60 & 3 &  05$^{\rm h}$27$^{\rm m}$10\farchs89& $-$67\deg27\arcmin27\farcs3& ub0h030 & F300W & $1 \times ~10$, $6 \times 160$ & 2008 Jul 18\\
             &    &                                                                       &                                                         &                 & F450W & $1 \times ~10$, $6 \times 160$ & 2008 Jul 18\\
             &    &                                                                       &                                                         &                 &  F555W & $4 \times 300$, $2 \times 350$ & 2008 Jul 18\\
             &    &                                                                       &                                                         &                 & F656N & $1 \times 300$, $1 \times 260$, $1 \times 800$ & 2008 Jul 18\\
             &    &                                                                       &                                                         &                 & F814W & $4 \times 260$, $2 \times 300$ & 2008 Jul 18\\
             & 4 &  05$^{\rm h}$27$^{\rm m}$28\farchs56& $-$67\deg27\arcmin19\farcs4& ub0h040 & F555W & $4 \times 300$, $2 \times 350$ & 2008 Jul 19\\                     
             &   &                                                                       &                                                         &                 & F814W & $4 \times 260$, $2 \times 300$ & 2008 Jul 19\\
LH~72 & 5 &  05$^{\rm h}$32$^{\rm m}$26\farchs03& $-$66\deg28\arcmin15\farcs8& ub0h050 & F300W & $1 \times ~10$, $6 \times 160$ & 2008 Jul 21\\
             &    &                                                                       &                                                         &                 & F450W & $1 \times ~10$, $6 \times 160$ & 2008 Jul 21\\
             &    &                                                                       &                                                         &                 &  F555W & $4 \times 300$, $2 \times 350$ & 2008 Jul 21\\
             &    &                                                                       &                                                         &                 & F656N & $1 \times 300$, $1 \times 260$, $1 \times 800$ & 2008 Jul 21\\
             &    &                                                                       &                                                         &                 & F814W & $4 \times 260$, $2 \times 300$ & 2008 Jul 21\\
             & 6 &  05$^{\rm h}$32$^{\rm m}$08\farchs59& $-$66\deg28\arcmin24\farcs0& ub0h060 & F555W & $4 \times 300$, $2 \times 350$ & 2008 Jul 14\\                     
             &   &                                                                       &                                                         &                 & F814W & $4 \times 260$, $2 \times 300$ & 2008 Jul 14\\
LH~88 & 7 &  05$^{\rm h}$35$^{\rm m}$55\farchs13& $-$67\deg34\arcmin39\farcs9& ub0h570 & F300W & $1 \times ~10$, $6 \times 160$ & 2008 Sep 1\\
             &    &                                                                       &                                                         &                 & F450W & $1 \times ~10$, $6 \times 160$ & 2008 Sep 1\\
             &    &                                                                       &                                                         &                 &  F555W & $4 \times 300$, $2 \times 350$ & 2008 Sep 1\\
             &    &                                                                       &                                                         &                 & F656N & $1 \times 300$, $1 \times 260$, $1 \times 800$ & 2008 Sep 1\\
             &    &                                                                       &                                                         &                 & F814W & $4 \times 260$, $2 \times 300$ & 2008 Sep 1\\
             & 8 &  05$^{\rm h}$35$^{\rm m}$37\farchs72& $-$67\deg34\arcmin50\farcs5& ub0h080 & F555W & $4 \times 300$, $2 \times 350$ & 2008 Jul 9\\                     
             &   &                                                                       &                                                        &                  & F814W  & $4 \times 260$, $2 \times 300$ & 2008 Jul 9  \\          
Field~1&  &  05$^{\rm h}$41$^{\rm m}$38\farchs52& $-$68\deg17\arcmin01\farcs8& u9e5550 & F555W  & $2 \times 500$& 2008 Feb 21\\
             &    &                                                                       &                                                         &                 & F814W  & $2 \times 500$& 2008 Feb 21\\
Field~2 &  &  05$^{\rm h}$24$^{\rm m}$04\farchs72& $-$67\deg21\arcmin57\farcs7& u9px070 & F555W  & $2 \times 500$& 2007 May 20\\
             &    &                                                                       &                                                         &                 & F814W  & $2 \times 500$& 2007 May 20
\enddata
%\tablenotetext{a}{}\tablenotemark{a}
\tablecomments{Observations of Fields~1 and 2, for the assessment of the general LMC field population, 
are observed within {\sl HST} programs GO-10583 (PI: C. Stubbs) and GO-10903 (PI: A. Rest) respectively. These data were retrieved from the {\sl HST Data Archive}.}
\end{deluxetable*}
%%%%%%%%%%%%%%%%%%%%%%%%%%%%%%%%%%%%%%%%%%%%%%%%%%%%%%%%%%%

The existence of such stars in star-forming regions of the LMC became recently known thanks to the 
angular resolution and wide-field coverage provided by {\sl Hubble Space Telescope} ({\sl HST}). Archival
images taken with the {\sl Wide-Field Planetary Camera 2} (WFPC2) of the young LMC association LH~52 
\citep{lucke70}, located at the northeastern edge of LMC~4, revealed 
for the first  time that low-mass PMS stars can be directly identified in the color-magnitude diagram 
(CMD) from photometry in $V$- and $I$-equivalent filters \citep{gouliermis06}. While the WFPC2 
images of LH~52 provided the first proof of the existence of such young stars in LMC star-forming regions, 
they were not deep enough to allow a statistically sound investigation of these stars. Subsequent deep 
photometry with the {\sl Advanced Camera for Surveys} (ACS) in the same filters of the association 
LH~95, located at the northwest periphery of LMC~4, revealed an outstanding sample of more than 
2,500 low-mass PMS stars down to the (model-dependent) mass-limit of $\sim$~0.2~M{\solar}, the smallest 
stellar mass ever observed in another galaxy \citep{gouliermis07}. 

%\clearpage 
%%%%%%%%%%%%%%%%%%%%%%%%%%%% FIGURE %%%%%%%%%%%%%
\begin{figure*}[t!]
\centerline{\includegraphics[angle=0,clip=true,width=\textwidth]{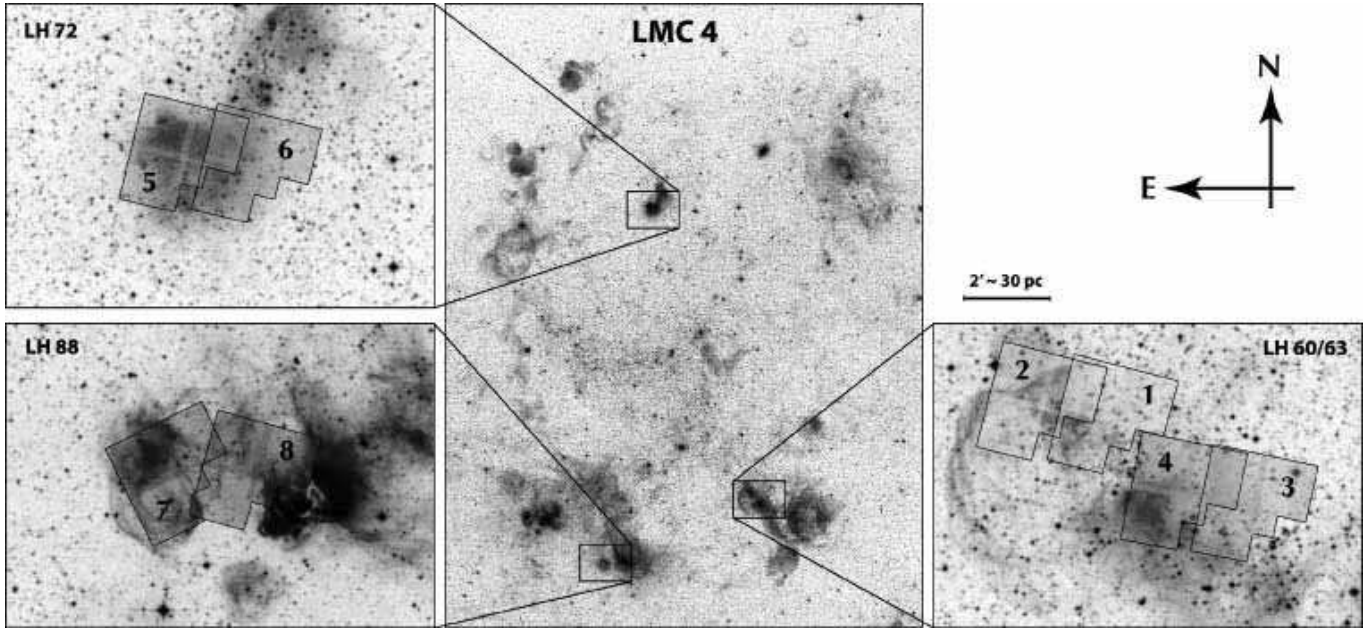}}
\caption{Maps of the regions observed. The general area of the super-giant shell LMC~4 is shown in the center 
from a scanned UK Schmidt telescope plate observed in the $R$ band, available from the DSS. The
extracted subregions, shown on the left and right of the figure, cover about 10\arcmin~$\times$~7\arcmin\ 
around the targets. The observed WFPC2 pointings are overlaid and annotated according to the
corresponding visit number (see also Table~\ref{t:obs}). The scale indicator corresponds to the 
fields-of-view of the individual regions around the observed clusters. The field-of-view of the image 
of the general LMC~4 area is $\sim$~1.7\deg~$\times$~2.2\deg\ (about 1.5~kpc~$\times$~2~kpc). 
\label{f:targetmap}}
\end{figure*}
%%%%%%%%%%%%%%%%%%%%%%%%%%%%%%%%%%%%%%%%%%%%%%

{ These data are characterized by an unprecedented completeness in their photometry due to their deepness. }
They provided us, thus,  a unique sample of PMS stars, which we utilized to address in detail the formation 
of a young stellar cluster in another galaxy. Specifically, we addressed the initial mass function (IMF) of LH~95 
down to the sub-solar regime \citep[][]{dario09} and the age of the cluster, as well as the duration of the star formation 
process in it, via the detection of an age-spread among its low-mass PMS stars \citep[][]{dario10}. 
The aforementioned investigations provide a unique insight of the faintest stars in two star-forming
regions within LMC~4. However, a more complete analysis of PMS populations in this area  
requires the collection of data on a larger sample of such objects. Within our {\sl HST} program GO-11547 
we obtained multi-band imaging of four additional star-forming regions located at the periphery of 
LMC~4 in order to accomplish a thorough characterization of PMS, MS and evolved stars in their vicinity, 
study their formation process, determine the stellar IMF and its variations in different environments, 
and provide an ample set of physical parameters of all observed stellar types in these regions. 

In this first paper we present the photometry, identify the low-mass PMS stellar
 populations in the observed regions, and determine their cluster membership via their CMD positions. Specifically, in \S~\ref{s:phot} we present the observations of this program and describe the 
photometric process, as well as its accuracy and completeness. Our measurements of the interstellar 
reddening towards the targets, based on the photometry of the brighter MS stars, are presented in 
\S~\ref{s:red}. We present the observed CMDs in \S~\ref{s:cmd}, where we also discuss the CMD of 
the general LMC field. The identification { and a qualitative assessment} of the PMS stars in the 
observed regions takes place in \S~\ref{s:pmsvar}, where the Hess diagrams and a {\sl Monte Carlo} 
technique for the statistical subtraction of field stars from the observed CMDs are described. 
In \S~\ref{s:pmsmmb} we present the quantitative determination of the true PMS populations of the 
observed regions from the stellar distributions along cross-sections of the observed CMDs. We 
discuss our findings in \S~\ref{s:disc}. Concluding remarks and our plans for subsequent analyses 
are given in \S~\ref{s:concl}. 

\section{Observations and Photometry} \label{s:phot}

\subsection{Observations}

Our investigation is based on the analysis of data collected within our {\sl HST} program 
GO-11547 (PI: D. Gouliermis) that imaged young stellar clusters located at the apparent 
edge of the super-giant shell LMC~4. Our targets are the young associations LH~60, LH~63, LH~72 
and LH~88 \citep{lucke70}, embedded in the bright emission nebulae DEM~L~201, 228 and 241 
\citep{dem76}, coinciding with the {\sc Hii} regions LHA 120-N~51A,C, N~55A and N~59C
\citep{henize56}, respectively. Maps of the general region of LMC~4 and the immediate areas 
of the targets are shown in Fig.~\ref{f:targetmap}.  
As such star-forming regions extend to sizes up to a few 10 pc (1\arcmin\ $\simeq$ 
15 pc at the distance of the LMC), our investigation requires the best combination of 
high angular resolution, to overcome crowding effects for faint stars, and wide 
field-of-view to effectively sample each region. During Cycle 16 the most appropriate 
instrument for this task was WFPC2. To achieve 
large spatial coverage each cluster was observed with two overlapping pointings. The first pointing
was centered on the brightest part of the associated {\sc Hii} region and WFPC2 imaging was 
performed in the broad-band filters F300W, F450W, F555W, F814W and the narrow-band 
filter F656N, roughly equivalent to standard $U$, $B$, $V$, $I$ and H\alp, respectively. We  
refer to each of these first group of multi-band pointings {\sl on} the {\sc Hii} regions as {\sl Pointing 1}. 
The second 
WFPC2 pointings of each cluster, referred to as {\sl Pointing 2}, are centered {\sl off} the 
bright nebula and were observed only in the F555W and F814W  filters. 

%located at the periphery of LMC~4

Since we wish to eliminate confusion effects, especially for the detection of the low-mass 
PMS populations, all observations are performed with a suitable dithering, using a sub-pixel 
pointing pattern, in order to improve the spatial resolution.  Furthermore, long exposures 
were split into 
intervals to aid in the removal of cosmic rays without any significant cost in the $S/N$ of the 
final combined image. Short exposures in the F300W  and F450W filters are used to obtain 
photometry for the brightest and youngest stars in these regions. Previous ground-based 
photometry is also available for these stars \citep[e.g.,][]{hill94, olsen01}. The consideration of 
the guide star acquisition, reacquisition of the telescope, and the instrument overhead times 
(e.g., change of filter, CCD readout, dithering) resulted in one set of four orbits for each
Pointing 1 and two orbits for each Pointing 2. The orbits for each pointing were grouped as 
a visit, for a total of 8 visits using a total of 24 orbits (Fig.~\ref{f:targetpix}). A detailed summary
of the observations is given in Table~\ref{t:obs}. For the assessment of the stellar populations 
in the general field of the LMC in the vicinity of the observed systems we used WFPC2 observations of the
LMC field available in the {\sl HST Data Archive}. We retrieved and analyzed images obtained in two 
campaigns, observed with parameters similar to ours. Details about these data are given also in 
Table~\ref{t:obs}. 

%%%%  NICMOS %%%%%
%Parallel observations with the {\sl Near Infrared Camera and Multi-Object 
%Spectrometer} (NICMOS), and specifically camera NIC2 in the F110W, 
%F160W and F205W ($J$-, $H$- and $K$-equivalent) filters are also performed.  

%These exposures will allow us, with a slightly extra charge of time, to characterize the
%faint, red, low-mass population of the surrounding LMC field.

%WFPC2, which with the use of a suitable dithering strategy can provide higher spatial resolution

%%%%%%%%%%%%%%%%%%%%%%%%%%%%%%%%%%%%%%%%%%%%%%%%%%%%%%%%%%%%%%%%%%%%%%%%
\begin{figure*}[t!]
%\centerline{\includegraphics[width=\textwidth]{f2rr.eps}}
\vspace*{1.truecm}
\centerline{\Large Figure is omitted due to size limitations.} 
\centerline{\Large Copy available upon request to the first author.}
\vspace*{1.truecm}
\caption{Color composite images of the primary WFPC2 observations of our program GO-11547 
constructed from the reduced images taken in the filters  F555W (blue) and F814W (red). 
Top panel shows all Pointings 1, i.e., visits 1, 3, 5, and 7. Bottom panel shows the
Pointings 2, i.e., visits 2, 4, 6, and 8. The corresponding cluster names are 
also given.  \label{f:targetpix}}
\end{figure*}
%%%%%%%%%%%%%%%%%%%%%%%%%%%%%%%%%%%%%%%%%%%%%%%%%%%%%%%%%%%%%%%%%%%%%%%%

%\clearpage
%%%%%%%%%%%%%%%%%%%%%%%%%%%% TABLE %%%%%%%%%%%%%%%%%%%%%%%%%%%%%%%%%%%%%%% 
\begin{deluxetable}{lrrrrr}
\tablecolumns{6}
\tablewidth{0pc}
\tabletypesize{\scriptsize}
%%\tablenum{1}
\tablecaption{Statistics on stars detected in filter-pairs with our photometry in each of Pointings 1. \label{t:phtstat} }
\tablehead{
\colhead{} &
\colhead{F300W} &
\colhead{F450W} &
\colhead{F555W} &
\colhead{F814W} &
\colhead{F656N} 
} 
\startdata
%\multicolumn{6}{c}{LH~63}\\
\cutinhead{LH~63}
F300W &     579 &     565 &     453 &     515 &     418 \\
F450W &         &    2~538 &    2~389 &    2~464 &     500 \\
F555W &         &         &    5~027 &    4~869 &     394 \\
F814W &         &         &         &    5~753 &     452 \\
F656N &         &         &         &         &     559 \\
%\multicolumn{6}{c}{LH~60}\\
\cutinhead{LH~60}
F300W &     348 &     340 &     294 &     316 &     257 \\
F450W &         &    2~509 &    2~436 &    2~467 &     357 \\
F555W &         &         &    5~237 &    5~050 &     309 \\
F814W &         &         &         &    5~551 &     332 \\
F656N &         &         &         &         &     409 \\
%\multicolumn{6}{c}{LH~72}\\
\cutinhead{LH~72}
F300W &     445 &     435 &     362 &     401 &     282 \\
F450W &         &    2~022 &    1~920 &    1~963 &     344 \\
F555W &         &         &    3~801 &    3~698 &     264 \\
F814W &         &         &         &    4~493 &     306 \\
F656N &         &         &         &         &     421 \\
%\multicolumn{6}{c}{LH~88}\\
\cutinhead{LH~88}
F300W &     206 &     201 &     173 &     180 &     141 \\
F450W &         &    1~777 &    1~719 &    1~737 &     229 \\
F555W &         &         &    3~819 &    3~715 &     201 \\
F814W &         &         &         &    4~321 &     207 \\
F656N &         &         &         &         &     285 
\enddata
\end{deluxetable}
%%%%%%%%%%%%%%%%%%%%%%%%%%%%%%%%%%%%%%%%%%%%%%%%%%%%%%%%%%%%

\subsection{Photometry}

Photometry is performed using the package HSTphot (Ver. 1.1) specifically designed for 
WFPC2 imaging \citep[][]{dolphin00a}. HSTphot is tailored to handle the undersampled nature of the
point-spread function (PSF) in WFPC2 images and uses a self-consistent
treatment of the charge transfer efficiency (CTE) and zero-point photometric calibrations. 
This package has the ability to perform photometry simultaneously for all exposures in 
different filters. Another advantage of HSTphot is that it allows the use
of PSFs which are computed directly to reproduce the shape details of star
images as obtained in the different areas of WFPC2 chips.  For this reason, we
adopt the PSF fitting option in the photometry routine, rather than use
aperture photometry. Photometric calibrations and transformations were made 
according to \cite{dolphin00a}, and CTE corrections were made according to \cite{dolphin02}. 
We used the HSTphot subroutine {\em mask} to take
advantage of the data quality files by removing bad columns
and pixels, charge traps and saturated pixels.  The same procedure is also
able to properly solve the problem of the vignetted regions at the chip edges. 
The subroutine {\em crmask} was used for the removal of cosmic rays.  

The HSTphot photometry routine, {\em hstphot},
returns data quality parameters for each detected source, which
can be used for the removal of spurious objects. After removing bad detections 
based on the quality parameters, we derived the final photometric catalogs
of all stars found in at least two of the considered filters. The numbers of stellar detections
with good photometric quality for each of the WFPC2 Pointings 1 are given in 
Table~\ref{t:phtstat}, paired in all observed filters. From these numbers 
it can be seen that the richest stellar samples are derived in the filters F555W and F814W
($V$- and $I$-equivalent). This is due to the higher sensitivity of the camera in the
corresponding wavelengths in combination with the applied long exposure times in these
filters. As far as Pointings 2 are concerned, observed only in these two filters, the 
numbers of stars detected in both filters with high photometric accuracy are 4~397 in LH~63, 
5~094 in LH~60, 3~886 in LH~72, and 3~958 in LH~88.

%\clearpage
%%%%%%%%%%%%%%%%%%%%%%%%%%%% TABLE %%%%%%%%%%%%%%%%%%%%%%%%%%%%%%%%%%%%%%% 
\begin{deluxetable*}{rccrrcccccccccc}
\tablecolumns{6}
\tablewidth{0pc}
\tabletypesize{\footnotesize}
%\rotate%%\tablenum{1}
\tablecaption{Sample of the multi-band photometric catalog of stars primarily detected in both F555W and F814W filters.  \label{t:phot} }
\tablehead{
\colhead{Star} &
\colhead{R.A.} &
\colhead{Decl.} &
\colhead{X} &
\colhead{Y} &
\colhead{$m_{\rm 300}$} &
\colhead{$\epsilon_{\rm 300}$} &
\colhead{$m_{\rm 450}$} &
\colhead{$\epsilon_{\rm 450}$} &
\colhead{$m_{\rm 555}$} &
\colhead{$\epsilon_{\rm 555}$} &
\colhead{$m_{\rm 814}$} &
\colhead{$\epsilon_{\rm 814}$} &
\colhead{$m_{\rm 656}$} &
\colhead{$\epsilon_{\rm 656}$} \\
\colhead{ \#} &
\multicolumn{2}{c}{(deg J2000)}&
\multicolumn{2}{c}{(pixels)}&
\colhead{} &
\colhead{} &
\colhead{} &
\colhead{} &
\colhead{} &
\colhead{} &
\colhead{} &
\colhead{} &
\colhead{} &
\colhead{} 
} 
\startdata
     1   &81.9437485 & $-$67.4254913  &149.25   &562.03 & 20.489 &0.127 &19.280 &0.007 &18.407 &0.002 &17.205 &0.002 &17.361 &0.022\\
     2   &81.9597015  &$-$67.4308624  &642.66   &148.52  &17.245 &0.008& 18.327& 0.003 &18.381 &0.002 &18.437 &0.003 &18.274 &0.030\\
     3   &81.9557800 & $-$67.4288330  &466.22  & 241.60 & 99.999 &9.999 &20.181 &0.007 &19.108 &0.002 &17.397 &0.002 &17.737 &0.017\\
     4   &81.9397507 &$-$67.4293671  &432.98   &729.03  &17.231 &0.009 &18.404 &0.004 &18.472 &0.002 &18.565 &0.003 &18.171 &0.033\\
     5   &81.9583130 & $-$67.4300308 & 571.80   &180.04 & 17.182& 0.006 &18.321& 0.003 &18.402 &0.002 &18.477 &0.003 &18.214 &0.025\\
     6   &81.9457169  &$-$67.4278641 & 343.66  & 531.86 & 17.543 &0.008 &18.570& 0.003 &18.537 &0.002 &18.368 &0.003 &18.164 &0.026\\
     7   &81.9394455  &$-$67.4292374  &421.46 &  736.55  &17.805 &0.012 &18.637 &0.005 &18.623 &0.002 &18.643 &0.003 &18.251 &0.042\\
     8   &81.9534912 & $-$67.4265442 & 276.55  & 282.40 & 17.527 &0.009 &18.653 &0.003 &18.727 &0.002 &18.775 &0.003 &18.486 &0.028\\
     9   &81.9536591 & $-$67.4270477 & 316.44  & 283.45 & 18.109 &0.012 &18.827 &0.004 &18.852 &0.002 &18.818 &0.003 &18.609 &0.035\\
    10  & 81.9566727 & $-$67.4309464 & 635.36 &  240.53 & 20.855 &0.079 &19.856 &0.007 &19.213 &0.003 &18.185 &0.002 &18.255 &0.024\\
    11   &81.9431229 & $-$67.4300385 &501.62   &635.93  &18.301& 0.019 &19.129 &0.007 &19.098 &0.002 &19.022 &0.004 &18.708 &0.054\\
    12  & 81.9473572  &$-$67.4330444  &755.68  & 545.32 & 18.312 &0.019 &19.074 &0.005 &19.082 &0.003 &19.069 &0.004 &18.818 &0.056\\
    13   &81.9619293 & $-$67.4312439  &682.93  &  86.19  &17.757 &0.021 &18.759 &0.007 &18.822 &0.004 &18.882 &0.005 &18.577 &0.032\\
    14  & 81.9591370 & $-$67.4265823  &305.71  & 113.45 & 21.080 &0.072 &19.995 &0.013 &19.314 &0.004 &18.262 &0.003 &18.312 &0.024\\
    15   &81.9490433  &$-$67.4320602  &687.11  & 482.87 & 18.534 &0.016 &19.196 &0.004 &19.176 &0.002 &19.176 &0.004 &18.942 &0.058
\enddata
\tablecomments{In this example the 15 first records of the catalog for LH~63 is shown. `99.999 $\pm$ 9.999' values correspond to non-detection
of the source in the specific filter.}
\end{deluxetable*}
%%%%%%%%%%%%%%%%%%%%%%%%%%%%%%%%%%%%%%%%%%%%%%%%%%%%%%%%%%%%

%\clearpage
%%%%%%%%%%%%%%%%%%%%%%%%%%%%%%%%%%%%%%%%%%%%%%%%%%%%%%%%%%%%%%%%%%%%%%%%
\begin{figure*}[t!]
\centerline{\includegraphics[width=0.75\textwidth]{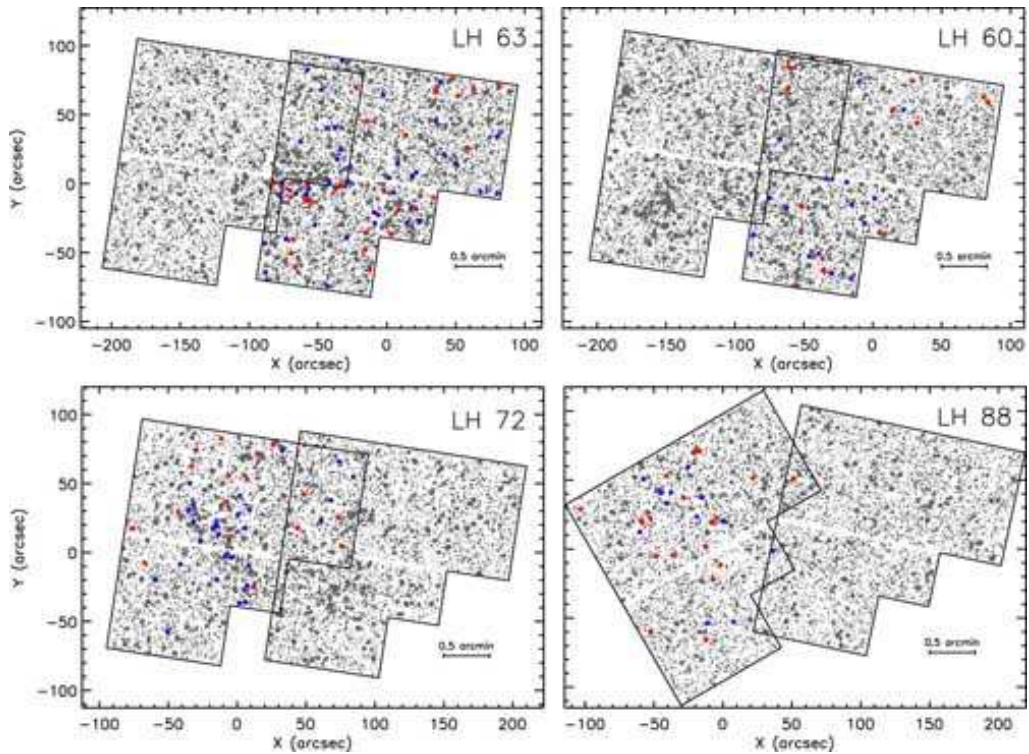}}
%\vspace*{1.truecm}
%\centerline{\Large Figure is omitted due to size limitations.} 
%\centerline{\Large Copy available upon request to the first author.}
%\vspace*{1.truecm}
\caption{Stellar charts of all stars detected with our WFPC2 photometry in both F555W and F814W
filters in each of the observed {\sc Hii} regions. Coordinates are given in seconds of arc from a reference 
point, which is selected to coincide with the center of WFPC2 Pointing 1 for every region. North 
is up and East is to the left of the charts. { Bright candidate accreting PMS stars, identified from their H\alp-excess 
(see \S~\ref{s:halpha}), and OB-type stars found with our photometry in shorter 
wavelengths (\S~\ref{s:cmd}) are marked with red and blue
symbols repsectively.}\label{f:charts}}
\end{figure*}
%%%%%%%%%%%%%%%%%%%%%%%%%%%%%%%%%%%%%%%%%%%%%%%%%%%%%%%%%%%%%%%%%%%%%%%%

%\clearpage
%%%%%%%%%%%%%%%%%%%%%%%%%%%% FIGURE %%%%%%%%%%%%%%%%%%%%%%%%%%%%%%%%%%%%%%% 
\begin{figure*}[t!]
%\centerline{\includegraphics[width=0.975\textwidth]{f4rr.eps}}
\vspace*{1.truecm}
\centerline{\Large Figure is omitted due to size limitations.} 
\centerline{\Large Copy available upon request to the first author.}
\vspace*{1.truecm}
\caption{Uncertainties of photometry as derived by HSTphot
for all filters in the fields covered by Pointings 1 for each cluster.\label{f:phterr}}
\end{figure*}
%%%%%%%%%%%%%%%%%%%%%%%%%%%%%%%%%%%%%%%%%%%%%%%%%%%%%%%%%%%%%%%%%%%%%%%%%%%

%\clearpage
%%%%%%%%%%%%%%%%%%%%%%%%%%%% FIGURE %%%%%%%%%%%%%%%%%%%%%%%%%%%%%%%%%%%%%%% 
\begin{figure*}[t!]
%\centerline{\includegraphics[width=0.975\textwidth]{f5rr.eps}}
\vspace*{1.truecm}
\centerline{\Large Figure is omitted due to size limitations.} 
\centerline{\Large Copy available upon request to the first author.}
\vspace*{1.truecm}
\caption{Completeness as it is evaluated by our artificial star experiments for the four broad-band filters
in the regions covered by each Pointing 1.\label{f:compl}}
\end{figure*}
%%%%%%%%%%%%%%%%%%%%%%%%%%%%%%%%%%%%%%%%%%%%%%%%%%%%%%%%%%%%%%%%%%%%%%%%%%%

We combined the photometric catalogs derived from the two pointings per cluster 
into one final catalog of stars detected in each targeted region. This
process revealed 12~511 stars in LH~60, 11~785 in LH~63, 9~756 in LH~72 and 
10~597 in LH~88 detected in both F555W and F814W filters. 
According to the methodology originally presented by \cite{gouliermis06}, and subsequently 
applied to various star-forming regions in both of the Magellanic Clouds 
\citep[see, e.g.,][]{sabbi07, gouliermis07, schmalzl08, cignoni09, vallenari10}, deep 
{\sl HST} imaging in $V$- and $I$-equivalent wavelengths 
provides an efficient detection of PMS stellar populations from their positions in the 
corresponding CMD. Therefore, we assess the PMS stars in the observed regions 
from the complete photometric catalogs of stars detected in both the F555W and F814W filters in both 
pointings of each cluster. The corresponding stellar charts of the regions are shown in Fig.~\ref{f:charts}. 
In addition, since these filters provide the richest stellar samples, we also base the final 
multi-band photometric catalog of each cluster on the catalog of sources detected in both the F555W and F814W
filters. A sample of this photometric catalog is shown in Table~\ref{t:phot}. Naturally, the photometry of stars identified in 
other pairs of filters are extremely useful and  will be considered in subsequent analyses. For example the samples of stars detected 
in both F300W and F656N ($U$- and H$\alpha$-equivalent) certainly host PMS stars with accretion, which 
will be revealed from their $U$- and H$\alpha$-excess emission \citep[see, e.g.,][]{romaniello04, guido2010}; 
an analysis that we will apply in a forthcoming study. Moreover,  multi-wavelength coverage of bright 
main-sequence (MS) stars is essential for the assessment of the interstellar reddening, which we perform 
later in this study.

\subsection{Photometric accuracy and completeness}\label{s:phterr-cmp}

Photometric uncertainty and incompleteness are two factors that strongly depend on the targeted region.
As a consequence both are higher in the frames centered on the {\sc Hii} nebulae, due to higher
stellar crowding and confusion by diffuse nebular emission. Fig. \ref{f:phterr} shows typical
uncertainties of photometry as a function of the magnitude for all filters, 
as obtained from our photometry on Pointings 1. As seen in Fig. \ref{f:phterr} , 
photometric uncertainties behave similarly for each region in the F450W, F555W, and F814W 
filters. The F555W and F814W observations reach much fainter magnitudes accurately than those 
with the F300W and F450W filters. On average, the uncertainties of our photometry have $\epsilon < $~0.1~mag 
for $m_{555}\lsim$~25.4 in all systems.

The 
completeness of our photometry  was evaluated on the basis of artificial star experiments performed with 
the native artificial star function built into HSTphot for the creation of simulated images by distributing artificial stars of 
known positions and magnitudes. This utility allows the distribution of stars with similar colors and
magnitudes as in the real CMD. { It should be mentioned that due to the small image size of artificial stars 
generated within HSTphot the appearance of few bright saturated stars in each observed region leads to a  
somewhat underestimated completeness.} The results of this process for the four broad-band filters 
used for Pointings 1 of each cluster are shown in Fig.~\ref{f:compl}. From these plots  it can be seen that 
our photometry in Pointing 1 is more complete systematically for stars found in filters F555W and F814W. 
In most of the clusters, the photometric completeness in the F300W filter drops at brighter magnitudes,  
$m_{\rm 300} \simeq$~22, while in F450W it is somewhat fainter. 
 On average, we reach the limit of 50\% completeness at $m_{\rm 555} \approx$~26. Completeness in both 
F555W and F814W in WFPC2 Pointings 2 is slightly improved due to lower crowding and confusion.

%\clearpage
%%%%%%%%%%%%%%%%%%%%%%%%%%%% FIGURE %%%%%%%%%%%%%%%%%%%%%%%%%%%%%%%%%%%%%%% 
\begin{figure*}[t!]
\centerline{\includegraphics[width=0.975\textwidth]{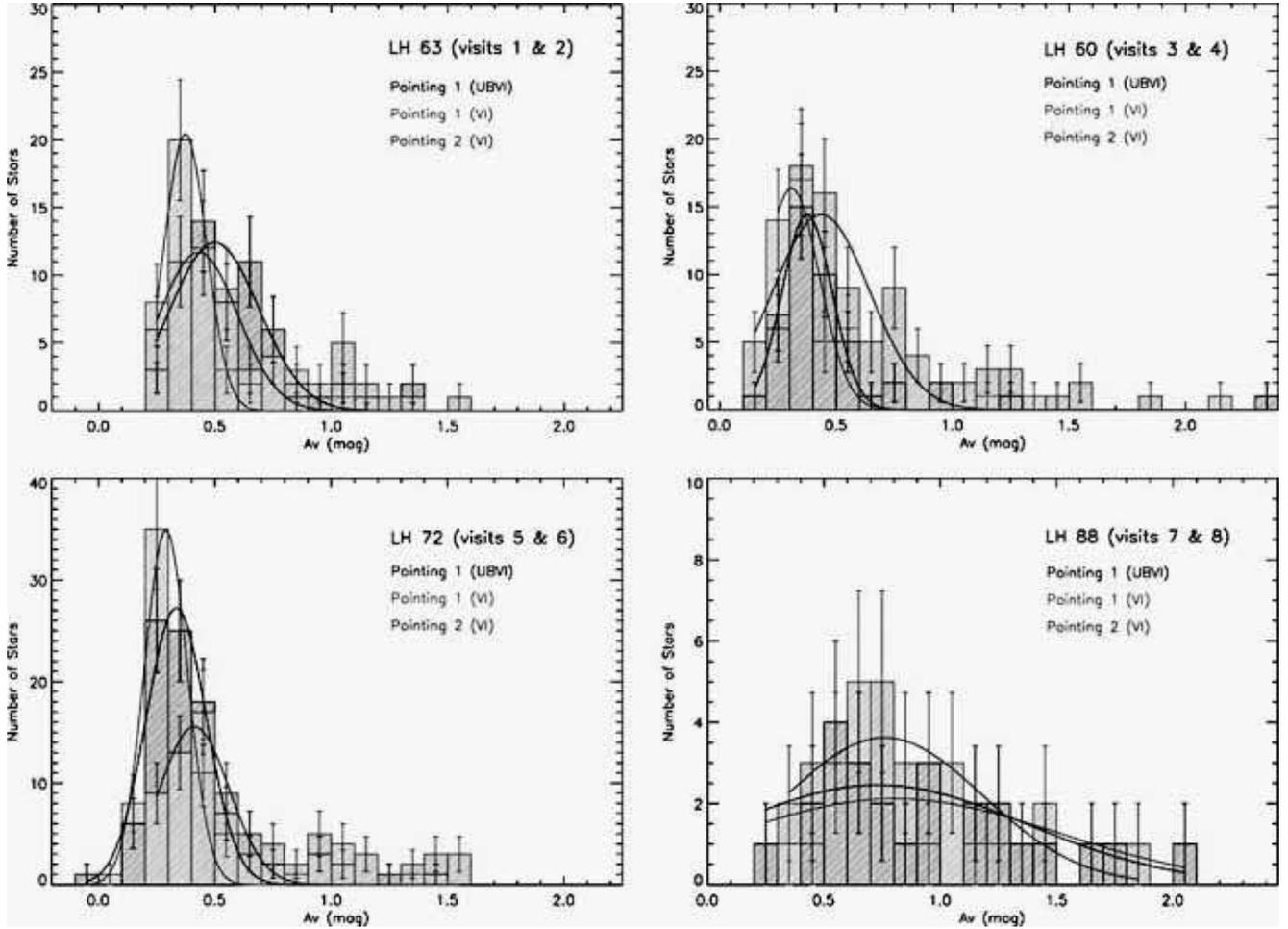}}
\caption{Visual extinction distributions as derived from upper MS stars with $S/N > 10$ with the use of 
observations in all four broad-band filters in Pointings 1, and in the $V$- and $I$-equivalents for
Pointings 2 for every region. We evaluate extinction in Pointings 1 also with the use of only $V$- and 
$I$-equivalent filters for reasons of completion and comparison to those for Pointings 2. 
The best-fitting distributions to the histograms are overlaid. The derived average (peak) $A_V$ values and
the corresponding 1$\sigma$ uncertainties are given in Table~\ref{t:av}. [A color version of this figure will be available in the online journal.] \label{f:av}}
\end{figure*}
%%%%%%%%%%%%%%%%%%%%%%%%%%%%%%%%%%%%%%%%%%%%%%%%%%%%%%%%%%%%%%%%%%%%%%%%%%%

\section{Interstellar Reddening}\label{s:red}

In this section we quantify the effect of interstellar extinction in the observed regions. We apply a statistical measurement 
of the visual extinction, $A_V$, 
in each observed region by constructing the distribution of the best-observed MS stars according to their extinction 
and obtaining its average value and 1$\sigma$ uncertainty. The calculation of $A_{V}$ toward 
the first WFPC2 pointing of each system is directly achieved from our 
observations in all four broad-band filters through comparison of data points in 
color-color diagrams to a model MS in the WFPC2 photometric system \citep{girardi02}.
For this calculation we considered only stars with $S/N > 10$ 
in all four bands, corresponding roughly to $m_{\rm 450} < 19.5$ and $m_{\rm 300} - m_{\rm 450} < 0$. 
However, the second WFPC2 pointing of each system is observed only in the bands F555W 
and F814W, and therefore we apply an additional independent  $A_{V}$ measurement based 
on the comparison of the observed CMD positions of the upper MS stars to that expected according to 
the model MS. This measurement is applied for MS stars with $S/N > 10$ in both bands, 
corresponding to $m_{\rm 555} < 19.5$ and $m_{\rm 555} - m_{\rm 814} < 0.5$.  

{ The extinction law depends on the nature and composition of 
interstellar dust grains. Therefore the law will vary depending on environment
such as Galactic star forming regions or for  
different galaxies like the Large and Small Magellanic Clouds 
\citep{gochermann02}.  Several investigations of the reddening 
law in the LMC show that the average LMC extinction curve does not differ
from that of the Milky Way at wavelengths covered by $B$-, $V$- and 
$I$-equivalent filters \citep[e.g.,][]{koornneef81, nandy81, fitzpatrick86, 
sauvage91, misselt99}, but with differences appearing in the $U$ 
band and for shorter wavelengths. Therefore, for the evaluation of 
extinction we assume the typical Galactic extinction law 
\cite[e.g.,][]{cardelli89, fitzpatrick90}, parameterized by a value of  
$R_V= A_V/E(B-V ) = 3.1$. Relative extinction was determined 
 in the WFPC2 passbands from the analytical expression of 
stellar extinction of \cite{fitzpatrick90}.  }

%\clearpage
%%%%%%%%%%%%%%%%%%%%%%%%%%%%%%%%%%%%%%%%%%%%%%%%%%%%%%%%%%%%
\begin{deluxetable}{cccc}
%\tabletypesize{\footnotesize}
\label{t:av}
\tablecolumns{4}
\tablewidth{0pc}
\tablecaption{Visual Extinction, $A_V$, toward the observed regions. \label{t:av} }
\tablehead{
\colhead{Region} & 
\multicolumn{2}{c}{WFPC2 Pointing 1}&
\colhead{WFPC2 Pointing 2} \\
\colhead{} & 
\colhead{``$UBVI$''} &
\colhead{``$VI$''} & 
\colhead{``$VI$''} 
} 
\startdata
LH 63& 0.49~$\pm$~0.19& 0.37~$\pm$~0.09 & 0.43~$\pm$~0.17\\
LH 60& 0.37~$\pm$~0.10 & 0.31~$\pm$~0.12 & 0.44~$\pm$~0.21 \\
LH 72& 0.33~$\pm$~0.13& 0.29~$\pm$~0.09& 0.41~$\pm$~0.15\\
LH 88& 0.73~$\pm$~0.64 & 0.81~$\pm$~0.70 & 0.76~$\pm$~0.42 
\enddata
%\tablenotetext{*}{}
\tablecomments{The given $\pm$ values are the measured spreads, $\sigma$, in $A_V$, not the uncertainties in the measured values.}
\end{deluxetable}
%%%%%%%%%%%%%%%%%%%%%%%%%%%%%%%%%%%%%%%%%%%%%%%%%%%%%%%%%%%

%\clearpage 
%systems_cmds-..v.i.+contour-maps.eps
%%%%%%%%%%%%%%%%%%%%%%%%%%%% FIGURE %%%%%%%%%%%%%
\begin{figure*}[t!]
\centerline{\includegraphics[clip=true,width=1.0\textwidth]{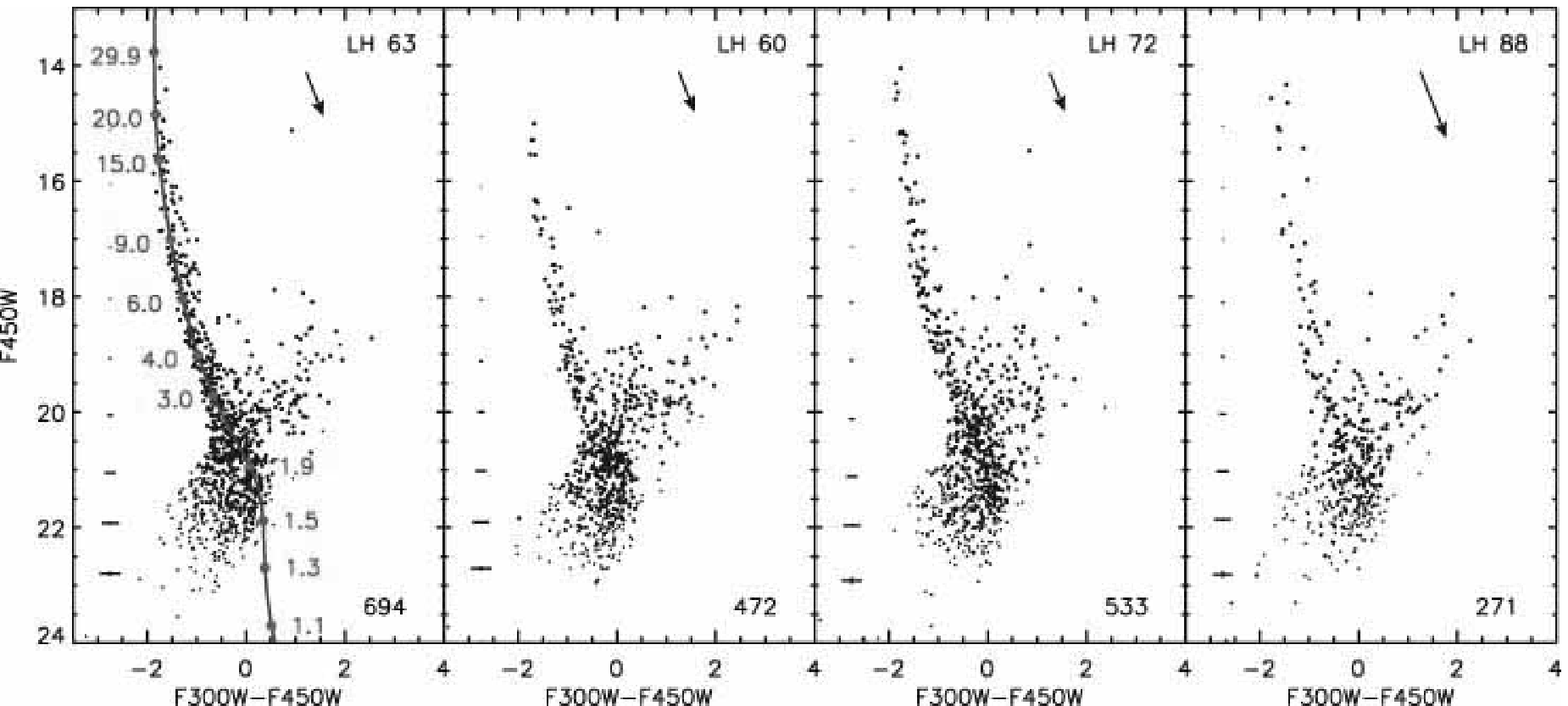}}
\caption{The  $UB$-equivalent CMDs of the stars detected with our photometry in both F300W and F450W 
filters in Pointings 1 of the observed regions. Arrows represent the maximum $A_V$ measurements
given in Table~\ref{t:av}. Typical photometric errors in both magnitudes and colors are shown on 
the left of each CMD. The numbers of stars with $\bar{\delta_2}\leq0.1$~mag are given at the bottom-right 
of each diagram. The positions of these stars 
are indicated in the  CMDs by thicker points. The ZAMS from the \cite{girardi02} 
grid of models is overlaid with indicative stellar mass values for MS stars. 
[A color version of this figure will be available in the online journal.] \label{f:ub-cmd}}
\end{figure*}
%%%%%%%%%%%%%%%%%%%%%%%%%%%%%%%%%%%%%%%%%%%%%%

Our results are shown in Fig.~\ref{f:av}, where we plot the derived $A_V$ distributions for each Pointing
of every observed region. The average extinction values per region and pointing are given with their
1$\sigma$ errors in Table~\ref{t:av}. The fits applied to 
the histograms of Fig.~\ref{f:av} consider {\sl all} stars within the selected $S/N$. { It should be noted, however, 
that the CMD positions
stars and most distant from the MS coincide with those
of emission-line star-disk systems such as  Be, B[e] stars \citep[e.g.,][]{wisniewski07, kraus09}, or  of
Herbig Ae/Be stars \citep[e.g.,][]{nishiyama07, clayton10}, which are expected in such star-forming regions.}
{ In addition, the appearance of binaries among MS stars will naturally 
lead to a small overestimation of the evaluated reddening.} 
As a consequence, the derived 
values and errors of $A_V$ should be considered as the {\sl upper limits} for the actual extinction. 
{ It is worth noting that, as shown from the values of Table~\ref{t:av}, while in general 
Pointing 2 is less crowded and less affected by the {\sc Hii} emission than Pointing 1, 
$A_V$ there is generally larger than in the latter. This is due to the higher concentration
of dust on the rim of the {\sc Hii} region, as is typical in such star-forming regions, which
produces higher extinction on the periphery of the nebula.}

In our subsequent analysis we will consider these upper values, rather than any lower average 
$A_V$ derived from the distributions after trimming the higher values bins. We will specifically 
use the maximum $A_V$ value, derived from the values of Table~\ref{t:av} as 
${\langle A_V \rangle}_{\rm max}+ \sigma_{\rm max}$  for each region. The reason for  this approach 
is the fact that low-mass PMS stars in star-forming regions 
of the Magellanic Clouds occupy the fainter-redder part of the visual CMD, which can be strongly affected by reddening  
\citep[see, e.g.,][]{gouliermis-eslab07}. As a consequence faint low- and intermediate-mass MS stars 
may be confused for low-mass PMS stars due to reddening. Therefore, since it is important to apply a safe 
distinction between faint reddened MS and true PMS stars we quantify the maximum possible effect of visual 
extinction to the observed stellar samples { in our qualitative distinction between PMS and 
 reddened MS stars in the Hess diagrams and field-subtracted CMDs of the observed regions 
 (see \S~\ref{s:pmsvar}). On the other hand, 
the quantitative selection of PMS stars according to their cross-section 
CMD distributions is not affected by the considerations of maximum 
reddening values, because these distributions in the CMDs of the
less affected regions LH~60, LH~63 and LH~72 peak at colors redder 
than those of the expected most-reddened MS stars according to our 
reddening determinations (see \S~\ref{s:pmsmmb}). }

%\clearpage 
%systems_cmds-..v.i.+contour-maps.eps
%%%%%%%%%%%%%%%%%%%%%%%%%%%% FIGURE %%%%%%%%%%%%%
\begin{figure*}[t!]
\centerline{\includegraphics[clip=true,width=1.0\textwidth]{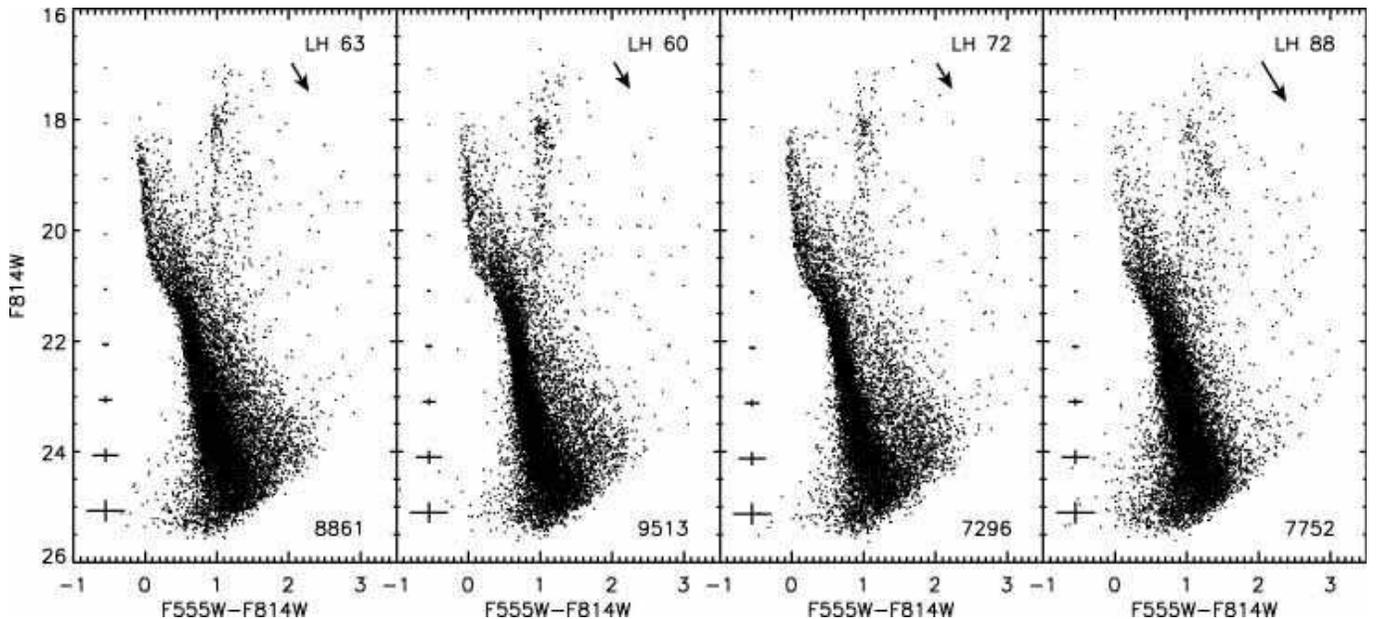}}
\caption{The  $VI$-equivalent CMDs of the stars detected with our photometry in both F555W and F814W 
filters in both Pointing 1 and 2 of each observed region. The numbers of stars with $\bar{\delta_2}\leq0.1$~mag 
are given at the bottom-right of each diagram. The loci
of the PMS stars cover the red sequence at $m_{\rm 814}$~\gsim~21, which is almost parallel to the MS. 
The intrinsic characteristics of these objects and possibly different ages among them are responsible for the 
spread of their CMD positions (see discussion in \S~\ref{s:disc}. It is interesting to note that the CMD of LH~88 
shows clear signs of extensive differential extinction in the system,  
apparent in the broad MS and elongated RC.\label{f:vi-cmd}}
\end{figure*}
%%%%%%%%%%%%%%%%%%%%%%%%%%%%%%%%%%%%%%%%%%%%%%

\section{Color-Magnitude Diagrams}\label{s:cmd}

The $m_{\rm 300}-m_{\rm 450}$ versus $m_{\rm 450}$ ($UB$-equivalent) and $m_{\rm 555}-m_{\rm 814}$ 
versus $m_{\rm 814}$ ($VI$-equivalent) CMDs of the stars detected in the 
areas of the clusters are plotted in Figs.~\ref{f:ub-cmd} and \ref{f:vi-cmd}, respectively. 
{ Since only Pointing 1 was observed in the $U$- and $B$- equivalent 
wavebands, the corresponding CMDs show the stellar populations included only in this pointing
of each cluster.
On the other hand, both Pointing 1 and 2 were observed in the $V$- and $I$- equivalent 
wavelengths, and therefore the corresponding CMDs show the stellar populations  
covered by both pointings.}
Reddening vectors corresponding to the highest $A_V$ values derived in 
\S~\ref{s:red}, as they are given in Table~\ref{t:av}, are also plotted in the CMDs. 
In order to select stars with the overall best photometry, we have used the average 
error in the two bands, $\bar{\delta_2}$,  considered for every CMD set, given as 
 \beq \displaystyle \bar{\delta_2} = \sqrt{\frac{\epsilon_{\rm 300}^2+\epsilon_{\rm 450}^2}{2}}
 ~~{\rm or}~~\bar{\delta_2} = \sqrt{\frac{\epsilon_{\rm 555}^2+\epsilon_{\rm 814}^2}{2}}~~,\eeq
 respectively. The numbers of stars with $\bar{\delta_2} \leq 0.1$~mag in every stellar 
 catalog are given at the bottom-right of each CMD. 
 
In the CMDs of Figs.~\ref{f:ub-cmd} and \ref{f:vi-cmd} it is shown that all observed regions 
are characterized by a variety of stellar species. The two sets of CMDs for every region, due 
to the selected filters and exposure times, track different types of stars, which nevertheless 
belong to the same system. In particular, the $UB$ CMDs of Fig.~\ref{f:ub-cmd} indicate a sharp 
upper main-sequence (UMS) populated by the blue intermediate- and high-mass stars, which 
typically characterize embedded young LMC associations. According the 
the ZAMS model (plotted over the CMD for LH~63) all observed clusters host well-populated
UMS mass functions with the most massive stars being O-type dwarfs with masses  
between 20 and 30~M{\solar}. These stars and their relation to the observed clusters will be 
the subject of another subsequent analysis of the massive stellar content of these clusters.
{ The OB stars revealed in the observed clusters from the $UB$-equivalent CMDs 
are shown with blue symbols in the stellar charts of Fig.~\ref{f:charts}.}

In the present analysis we concentrate on the rich $VI$ CMDs of the clusters (Fig.~\ref{f:vi-cmd}). 
These CMDs are populated by a larger variety of stellar sources, covering the MS (faint and bright), 
the turn-off (TO), the red clump (RC), the red giant branch  (RGB)  and the PMS. These 
features represent two separate populations in the LMC, i.e., the evolved population of 
the general galaxy-field and the young and currently forming stellar populations of the
clusters themselves. We describe each of these features in the following text. It should be noted 
that the 
observations presented in Fig.~\ref{f:vi-cmd} are mainly focused on the faint stars, 
and therefore long exposures were performed in these filters. As a consequence, 
the UMS and the tip of the RGB are saturated and thus not shown. All four CMDs of 
Fig.~\ref{f:vi-cmd} share all aforementioned features, and specifically apart from the 
UMS there is a pronounced TO at $m_{\rm 814} \simeq 21$ and a RGB with its RC located at 
$m_{\rm 814} \simeq 18$ and $m_{\rm 555}-m_{\rm 814} \simeq 1.0$. These two features are 
known to be typical of the LMC field, as we demonstrate later  (in \S~\ref{s:fdecont}; see also 
Fig.~\ref{f:fcmds}). Below the TO, 
moving to fainter magnitudes, the low main sequence (LMS) is the dominant feature
of the CMDs. Previous {\sl HST} imaging of other LMC star-forming regions  
has demonstrated that this population belongs {\sl entirely} to the general LMC field and not the
stellar systems \citep[][]{gouliermis07, vallenari10}. On the other hand, 
the faint stellar membership of the young clusters at roughly the same brightness range 
with the LMS is located at redder colors in the CMDs, where a prominent broad sequence 
of PMS stars can be clearly seen in three of the observed regions. As we show later 
(\S~\ref{s:pmsvar}) these 
stars belong {\sl only} to the star-forming regions, and they are {\sl not} related to the general 
LMC field.

%\clearpage
%%%%%%%%%%%%%%%%%%%%%%%%%%%% FIGURE %%%%%%%%%%%%%%%%%%%%%%%%%%%%%%%%%%%%%%%
\begin{figure*}[t!]
\centerline{\includegraphics[clip=true,width=1.0\textwidth]{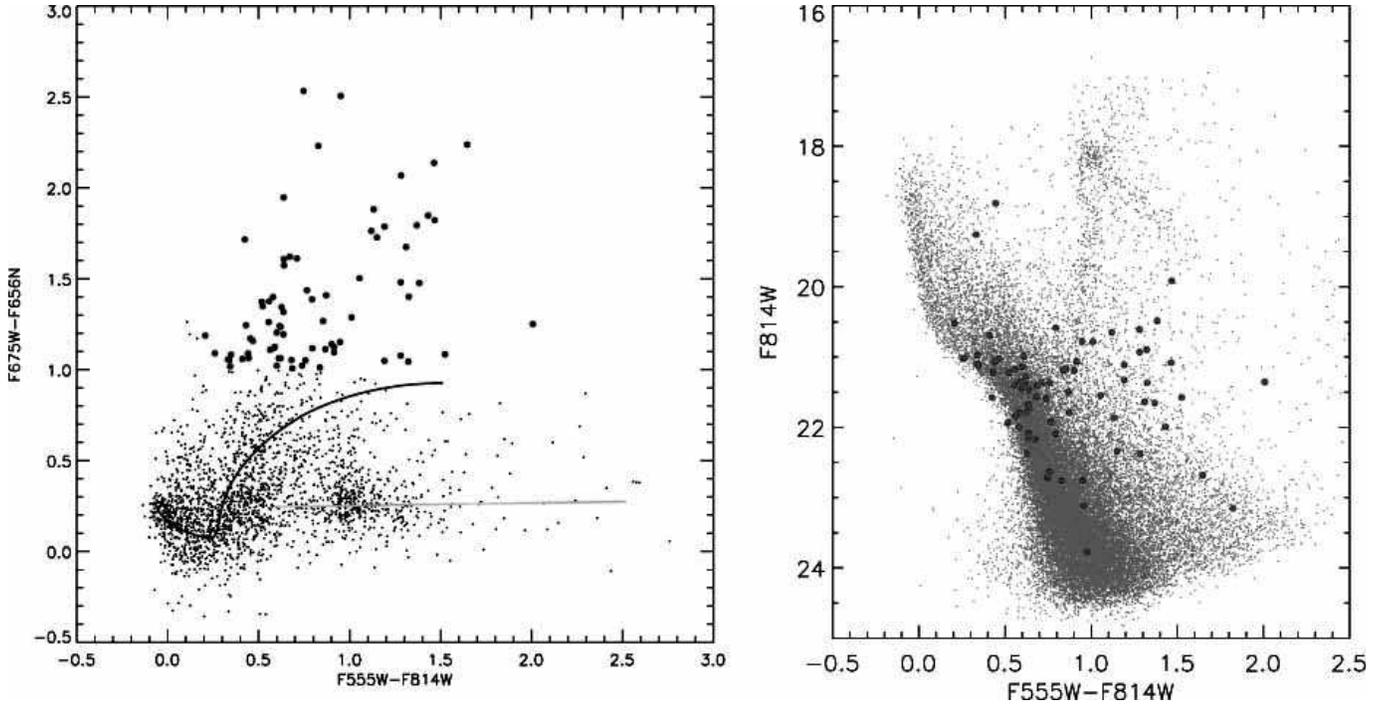}}
\caption{ {\em Left}: Diagnostic for stars with H{\alp} excess emission in the 
$R-{\rm H}\alpha$, $V-I$-equivalent color-color diagram. { In this diagram 
the positions of normal stars without  H\alp\ emission, i.e., field or weak-line 
T~Tauri stars, are indicated by the blue line, and reddened field stars by
the green. The group of stars with $R-{\rm H}\alpha \geq 1.0$~mag are mostly
stars, which exhibit H\alp\ excess (plotted with red symbols)}.  {\em Right}: The $VI$-equivalent 
CMD of the whole sample of observed stars in all four regions with 
H{\alp} excess emission candidate stars overlaid with red thick symbols.  
In this plot the CMD positions of bright PMS stars with H\alp\ excess can 
be readily identified, but the faint PMS populations are not covered due 
to the low sensitivity of the H\alp\ filter. The nature of the detected bright 
blue  sources, as well as of cool objects with H{\alp} 
excess will be investigated in a 
subsequent study. [A color version of this figure will be available in the online journal.]\label{f:haemit}}
\end{figure*}
%%%%%%%%%%%%%%%%%%%%%%%%%%%%%%%%%%%%%%%%%%%%%%%%%%%%%%%%%%%%%%%%%%%%%%%%%%%

%Bright sources with strong H\alp\ emission, which may be Herbig Ae/Be stars, or Be type stars,
%may also be found. 

\subsection{Stars with H\alp\ excess}\label{s:halpha}

PMS stars exhibit excess in their H{\alp} emission due to accretion. As a 
consequence the appearance of H{\alp} emission from stars in the clusters is a 
direct evidence of the existence of PMS stars in them. { Nevertheless, if the age of the observed 
clusters is of the order of $\sim$~3~Myr, H{\alp} emission stars should be very rare 
\citep[see, e.g., ][]{sung04b}. In addition, the accretion rate of most PMS stars in 
the clusters may be very low, and therefore only few active stars would show H{\alp} 
emission, because only stars with equivalent widths $W({\rm H}{\alpha}) > 10$~\AA\  
can be detected photometrically with confidence \citep{sung08}.}
Sources with H{\alp} excess 
emission can be photometrically identified by the comparison 
of their H\alp\ and $R$-equivalent magnitudes \citep[e.g.,][]{sung97}. 
As our dataset lacks observations in the $R$-equivalent WFPC2 
filter F675W, we determined a ``synthetic'' $m_{\rm 675}$ magnitude 
for every star derived by interpolation between data in the 
F555W and  F814W bands, as described by \cite{guido2010}.
In total, 2~263 stars are detected in all three F555W, F814W and 
F656N bands with the best photometric quality in all four observed 
regions. 
We identify the H\alp\ excess  stars in our 
sample by plotting the $m_{\rm 675}-m_{\rm 656}$ color index against 
the $m_{\rm 555}-m_{\rm 814}$ (Fig.~\ref{f:haemit}, {\em left}).
This diagnostic is equivalent to the use of the mean magnitude of 
$V$ and $I$ as a pseudocontinuum magnitude at H{\alp} \citep[see, 
e.g.,][]{sung00, sung04}. 

{ Considering that the reddening vector in Fig.~\ref{f:haemit} 
({\em left}) is nearly parallel to the $m_{\rm 555}-m_{\rm 814}$
axis, the clump of stars around $m_{\rm 555}-m_{\rm 814} \sim 1.0$ and
$m_{\rm 675}-m_{\rm 656} \sim 0.3$ represents highly reddened 
background stars. The green line in the figure 
indicates the meanline of these non-emission stars \citep[e.g.,][]{panagia00, sung04}.
In Fig.~\ref{f:haemit} ({\em left}) a large scatter in the $m_{\rm 675}-m_{\rm 656}$ 
index of normal stars, i.e.,  field stars or weak-line T~Tauri stars, can also be 
seen. Taking into account the positions of these stars, indicated  in the 
figure by the blue line, it is safe to select 
stars with $m_{\rm 675}-m_{\rm 656}$~\gsim~1.0 
as candidate H{\alp} emission stars.} As a consequence, 
for a first-order identification of the best 
candidates we apply tentatively the selection 
criterion requiring all H\alp\ emission stars to have $m_{\rm 
675}-m_{\rm 656} \geq 1.0$ mag. { Objects bluer 
than $m_{\rm 555}-m_{\rm 814} \simeq 0.2$ are probably Ae/Be stars
\citep{guido2010}, but considering the youthfulness of the clusters
we should also expect a few massive PMS stars, i.e., Herbig Ae/Be stars. }
This selection revealed 78 sources, which appear 
in red in the $m_{\rm 675}-m_{\rm 656}$, $m_{\rm 555}-m_{\rm 
814}$ diagram. These sources are marked with red symbols in the 
$VI$-equivalent CMD of all detected sources shown in Fig.~\ref{f:haemit}, 
({\em right}), and in the stellar maps shown in Fig.~\ref{f:charts}.  

While our F656N observations provide direct evidence that the observed 
clusters host accreting PMS stars of intermediate and high mass, they
are designed for the investigation of the bright stellar content of the 
clusters, and therefore they are not deep enough to cover the more
populous low-mass PMS stellar content of the clusters. In fact, the 
aforementioned process identified PMS accreting candidates down to 
only $m_{\rm 814} \simeq 22$, much brighter
than the vast majority of the low-mass PMS stars in the clusters. For the 
statistical identification of the latter we make use of the  rich 
$VI$-equivalent CMDs, as described in the following sections.
Nevertheless, a further thorough investigation of bright PMS stars with accretion, 
based on their H\alp\ equivalent widths, as well as their excess 
emission in the $U$-equivalent filter, and the 
analysis of the relation between H\alp\ strength and $U$-excess with  
accretion rate will be presented in a forthcoming paper.

\subsection{Contribution of the general LMC Field}\label{s:fdecont}

Naturally, the CMDs of Fig.~\ref{f:vi-cmd} are `contaminated' by the 
stellar populations of the general field of the LMC. In order to assess 
the field stellar population in the area of the observed regions, and evaluate this 
contamination, we retrieved from the 
{\sl HST Data Archive}  WFPC2 observations
of {\sl empty} regions of the LMC. Special care was taken to select observations, 
which are the most representative of the  LMC field in the general 
vicinity of our LMC~4 target regions. In addition the fields should preferably 
have negligible differential extinction, and 
observational setup comparable to our own. 
Two WFPC2 pointings covered in the F555W and F814W bands matched 
best these criteria. The fields are observed within {\sl HST} programs GO-10583 
(PI: C. Stubbs)  and GO-10903 (PI: A. Rest), both focused on the detection of 
LMC microlensing events in Cycles 14 and 15 respectively. Details on these 
observations are given in Table~\ref{t:obs}.  Both our clusters and the general field were 
observed with almost identical setup of the instrument and exposure times,  
allowing a direct comparison between the $VI$ CMDs of the clusters and the 
local LMC field.

\begin{figure}[t!]
\centerline{\includegraphics[clip=true,width=1.0\columnwidth]{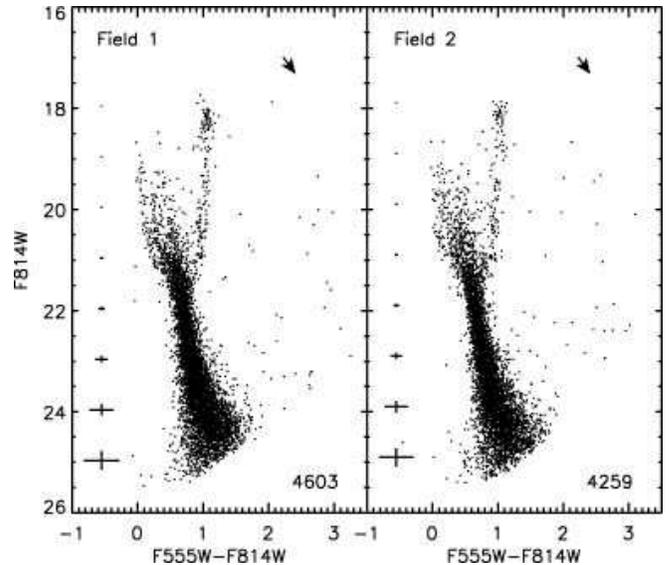}}
%\vspace*{1.truecm}
%\centerline{\large Figure is omitted due to size limitations.} 
%\centerline{\large Copy available upon request to the first author.}
%\vspace*{1.truecm}
\caption{The  $VI$-equivalent CMDs of the stars detected with our photometry in both F555W and 
F814W filters in the two selected empty areas for the assessment of the general field of LMC in the 
region of LMC~4. These WFPC2 images were retrieved from the {\sl HST Data Archive} 
(see \S~\ref{s:phot} and Table~\ref{t:obs}). The numbers of stars with $\bar{\delta_2}\leq0.1$~mag 
are given at the bottom-right of each diagram.  \label{f:fcmds}}
\end{figure}

The CMDs of these LMC control-fields, shown 
in Fig.~\ref{f:fcmds}, outline the complete star formation history (SFH) of the 
galaxy, which has been well constrained by earlier studies with WFPC2 imaging.
The results of these studies provide an accurate account of the 
evolutionary status of the stars we see in the CMDs of Fig.~\ref{f:fcmds}. 
According to these results, in general, the LMC disk is characterized by a continuous SFH for the 
last $\sim$~10 to 15~Gyr \citep[e.g., ][]{smecker-hane02}, with 
events of enhanced star formation that took place between 1 and 4~Gyr ago 
\citep{gallagher96, elson97, geha98, castro01, javiel05}. These events are
well recorded in the control-field CMDs of this study, as shown by the isochrones 
overlaid on the overall field CMD of Fig.~\ref{f:cmdsiso} ({\sl left}). These 
isochrones correspond to ages of roughly 0.8, 1.25, 2, 3, and 8~Gyr, proving 
this area to be somewhat younger than those described in the aforementioned 
studies, which are located outside LMC~4. A negligible visual extinction of 
$A_V=0.075$~mag, and the metallicity ($Z=0.008$) and distance of the LMC ($m-M = 18.5$~mag)
are applied to these models.

%%\clearpage 
%%systems_cmds-..v.i.+contour-maps.eps
%%%%%%%%%%%%%%%%%%%%%%%%%%%%% FIGURE %%%%%%%%%%%%%
%%%%%%%%%%%%%%%%%%%%%%%%%%%%%%%%%%%%%%%%%%%%%%%

\section{Identification of PMS Stars in The Observed Regions}\label{s:pmsvar}

\begin{figure}[]
\centerline{\includegraphics[clip=true,width=\columnwidth]{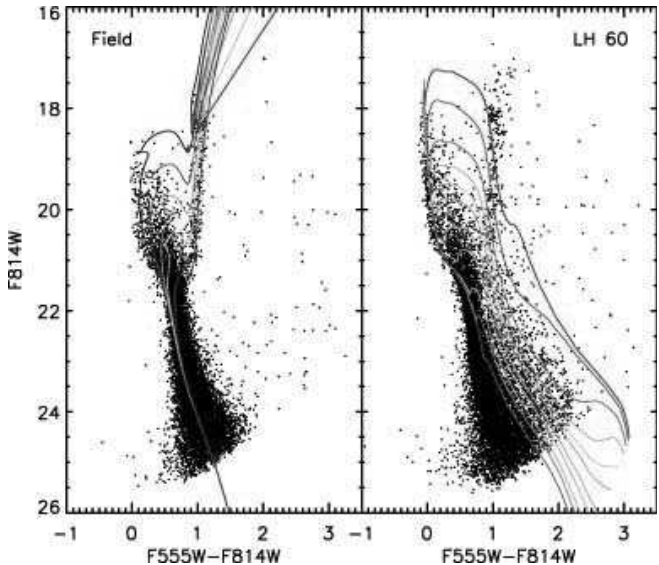}}
%\vspace*{1.truecm}
%\centerline{\large Figure is omitted due to size limitations.} 
%\centerline{\large Copy available upon request to the first author.}
%\vspace*{1.truecm}
\caption{{\sl Left}: $VI$-equivalent CMD of the stars in the general LMC field at the 
vicinity of LMC~4 with isochrones for evolved stars overlaid. The models are taken 
from the Padova grid \citep{girardi02} for the LMC metallicity ($Z=0.008$) and 
correspond to ages between roughly 1 and 10~Gyr in accordance with the 
established SFH of the galaxy. {\sl Right}: 
$VI$-equivalent CMD of the stars in the observed region of LH~60 with isochrones 
for PMS stars overlaid. The models are constructed with the {FRANEC} code in the 
WFPC2 photometric system and the metallicity of the LMC (see \S~\ref{s:pmsvar}).
[A color version of this figure will be available in the online journal.] \label{f:cmdsiso}}
\end{figure}

Low-mass PMS stars in LMC star-forming regions are known to cover the red-faint 
part of their $VI$-equivalent CMDs \citep[see, e.g.,][]{
vallenari10}, and the regions observed here do not seem to be an exception. 
Indeed, a visual comparison between the CMDs of Fig.~\ref{f:vi-cmd} and 
those of Fig.~\ref{f:fcmds} verifies the existence of a broad sequence of
PMS stars almost parallel to the LMS in the CMDs of the star-forming regions,
which is completely missing from the control-field CMDs. { In addition,
the reddening range of early-type stars in the star-forming 
regions (Table~\ref{t:av}), in comparison to that of the control-field ($A_V\simeq0.075$) 
shows that reddening alone cannot explain the existence 
of very red population almost parallel to the LMS in the CMDs of the observed
regions.} In this part of the analysis we
focus on methods for the identification of these PMS in the observed regions.
In the cases of LH\,60, LH\,63, and LH\,72 the appearance of
        PMS stars in the CMD is rather clear, but in the LH\,88 region the 
        higher reddening produces confusion in the unambiguous 
        identification of PMS stars. 
The radical difference in the comprised populations
between clusters and field is demonstrated more clearly in Fig.~\ref{f:cmdsiso},
where next to the overall control-field CMD with overlaid isochrones for evolved 
populations ({\sl left} panel), the CMD of LH~60 is plotted with overlaid 
isochrones for PMS populations ({\sl right} panel). These PMS models are constructed with 
the {\sl Frascati Raphson Newton Evolutionary Code} 
\citep[FRANEC,][]{chieffistraniero89,  deglinnocenti08} for the metallicity of the LMC
and the WFPC2 photometric system. They correspond to ages of 0.5, 1, 2, 3, 4, 
5, 7.5, 10, 15, 20 and 30~Myr. This broad age-coverage demonstrates a
well-documented problem in such observations, i.e., the wide spread 
of faint PMS stars, which does not necessarily correspond to any real 
age-spread among these stars, but it may well be the result of biases introduced
by observational constrains, { i.e., photometric accuracy and confusion,} and 
the physical characteristics of these stars, { such as variability, binarity and 
circumstellar extinction} \citep[see, e.g.,][for a detailed discussion]{dario10}.

%%\clearpage 
%%systems_cmds-..v.i.+contour-maps.eps
%%%%%%%%%%%%%%%%%%%%%%%%%%%%% FIGURE %%%%%%%%%%%%%
%%%%%%%%%%%%%%%%%%%%%%%%%%%%%%%%%%%%%%%%%%%%%%%

%%\clearpage 
%%systems_cmds-..v.i.+contour-maps.eps
%%%%%%%%%%%%%%%%%%%%%%%%%%%%% FIGURE %%%%%%%%%%%%%
\begin{figure*}[t!]
\centerline{\includegraphics[clip=true,width=1.\textwidth]{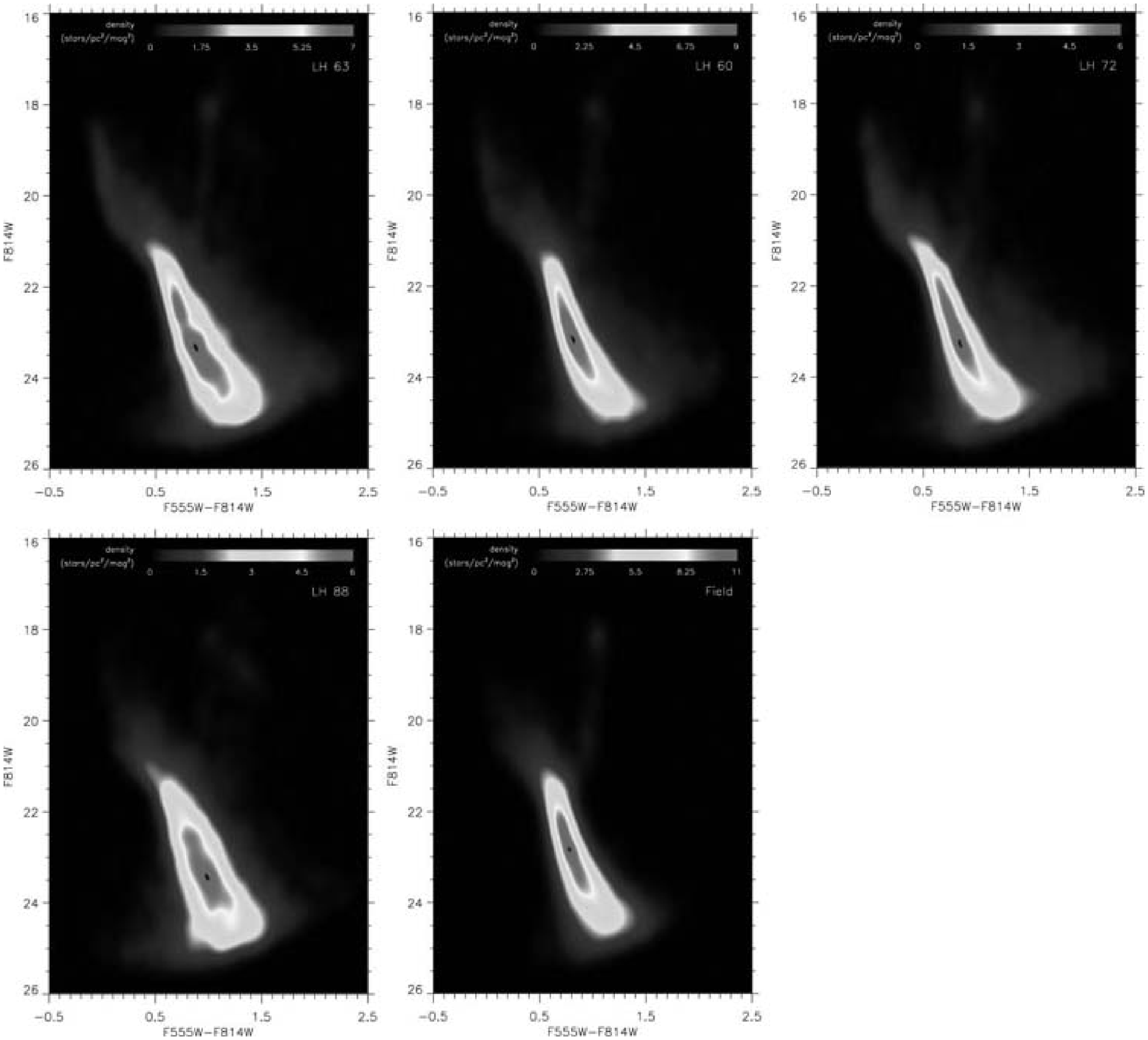}}
\caption{Hess diagrams of the observed regions, and the general LMC
field, constructed from their $VI$-equivalent CMDs. 
The diagrams are color-coded according to their 
stellar number densities indicated in the color bars. 
[A color version of this figure will be available in the online journal.] \label{f:hessor}}
\end{figure*}
%%%%%%%%%%%%%%%%%%%%%%%%%%%%%%%%%%%%%%%%%%%%%%%

The visual comparison of the CMDs of Fig.~\ref{f:cmdsiso} outline three important
aspects concerning the observed stellar populations: (1) The completely 
different evolutionary status of the field stars from that of the stellar content of  
the young clusters. (2) The great difficulty in distinguishing the evolved 
from the PMS populations at magnitudes brighter than $m_{\rm 814} \simeq 
21.5$. (3) At fainter magnitudes the heavy contamination of the cluster CMD by 
the field populations, particularly in the LMS. Taking into account these aspects,
it is important to establish an effective methodology for the determination of 
the {\sl true} PMS stellar population of each observed young cluster/star-forming region. 
In this section
we present such a scheme with the application of 
a qualitative approach based on the use of {\sl differential Hess Diagrams}, and of 
a statistical field-subtraction technique based on the {\sl Monte Carlo} method.
{ Although these techniques provide a first distinction of the redder
PMS  from the bluer MS low-mass population of the clusters, they are
indirect approaches in the determination of the true PMS populations. In particular,
with each of these methods one can assess  the field MS stellar contribution in the observed 
CMD in terms of stellar densities in the CMD, rather than on a star-by-star basis. 
Therefore, with these methods we cannot define the actual stars that should
be considered as the most probable PMS stars, except for the reddest ones.
As a consequence, we utilize these methods in order to demonstrate in a 
qualitative manner  the existence of low-mass PMS stars in the observed
regions.
A quantitative statistical determination of the actual membership of each 
PMS candidate will take place in the next section with the construction  
of the stellar distributions of the observed stars along {\sl cross-sections} 
of the faint part of each CMD. With this method a PMS membership probability
is assigned to each red star, according to its CMD position with respect to the 
cross-sections stellar distributions across the CMD.}
In our analysis we consider only stars with the
best photometric measurements, i.e., with $\bar{\delta_{2}} \leq 0.1$.

%%\clearpage 
%%systems_cmds-..v.i.+contour-maps.eps
%%%%%%%%%%%%%%%%%%%%%%%%%%%%% FIGURE %%%%%%%%%%%%%
\begin{figure*}[t!]
\centerline{\includegraphics[clip=true,width=1.\textwidth]{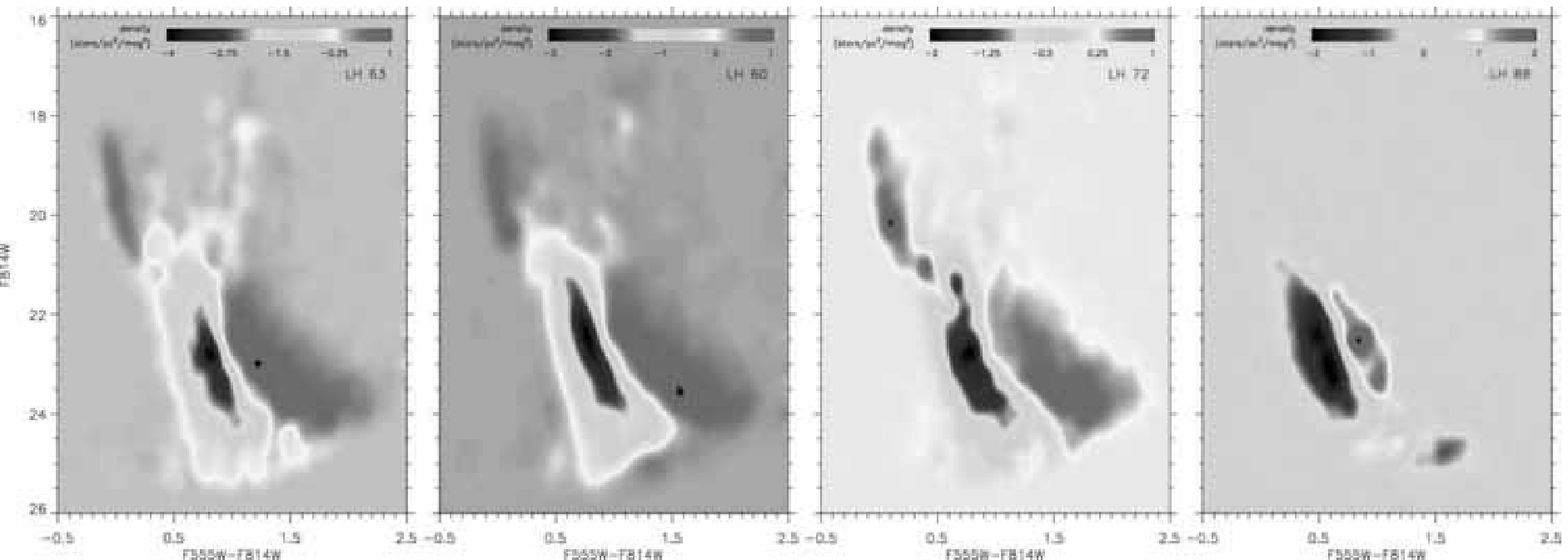}}
\caption{Differential Hess diagrams of the observed regions derived by 
the subtraction of the Hess diagram of the general LMC field from those of the 
observed regions. [A color version of this figure will be available in the online journal.]
\label{f:difhess}}
\end{figure*}
%%%%%%%%%%%%%%%%%%%%%%%%%%%%%%%%%%%%%%%%%%%%%%%

\subsection{Differential Hess Diagrams }\label{s:hessd}

Fig.~\ref{f:hessor}  shows the $VI$-equivalent Hess diagrams of the 
stars in the observed regions, as well as that of the overall general LMC 
field, i.e., the combined stellar content
of the two control-fields. { These diagrams are 
produced by binning the corresponding CMDs in 0.15 mag intervals in 
$m_{\rm 555} - m_{\rm 814}$ and 0.17~mag in $m_{\rm 814}$.  The 
size of the binning is selected to be small enough to reveal  
the fine-structure in the stellar density, but also large enough to construct a realistic 
representation of the observed CMDs. Very small bin sizes would produce 
`noisy' Hess diagrams, i.e., high numbers of small density peaks distributed 
throughout the diagrams, while large bin sizes would produce unrealistically 
thick Hess diagrams, which would smooth out any apparent effect of reddening 
or any true positional spread of stars in the CMD. The resulting two-dimensional 
distributions were  smoothed by interpolation with a boxcar average.} 
From the Hess diagrams of Fig.~\ref{f:hessor} it can be seen that 
the highest concentration of stars in all observed regions 
appears in the LMS. This part of the CMD is well-known to belong entirely to the general
LMC field (\S~\ref{s:fdecont}). In the cases of LH~60, LH~63 and LH~72, at redder colors
from the LMS and almost parallel to it,  
there is an apparent lower concentration of stars, which does not 
appear in the field. This feature corresponds to the faint PMS stars 
of the star-forming regions. 

The differences between the observed systems and the field are 
more apparent in the residual or differential Hess diagrams. These are
constructed with the subtraction of the field Hess diagram from those of the 
observed regions. { Before this subtraction, we simulated the CMD of the observed 
regions by shifting fraction of stars in the CMD of the control field according 
to the reddening law, so that the morphology of the RGB in the simulated CMD 
is similar to that of the observed ones. Then we adjusted the number 
of RGB stars of the simulated field CMD to that of the target region. In addition, 
a broadening of the MS band due to photometric error was also simulated 
using a Monte Carlo technique.} The resultant
differential Hess diagrams are shown in Fig.~\ref{f:difhess}. These diagrams 
for LH~60, LH~63 and LH~72 highlight the characteristic features of the 
$VI$-equivalent CMDs of typical LMC star-forming regions, i.e., a prominent UMS
down to the intermediate-mass regime and a broad sequence of red faint PMS 
stars reaching the sub-solar mass-regime.
{ In the differential Hess diagrams of Fig.~\ref{f:difhess} there are some  residual 
stellar concentrations remaining in particular at the RC and some locations of 
the RGB. They are most likely due to the small number of stars in the corresponding 
grids, which do not allow a proper subtraction from the CMD of the observed regions.
In addition, in the differential Hess diagrams of Fig.~\ref{f:difhess} there are no 
intermediate-mass PMS found in the Kelvin-Helmholtz contraction phase, located
between the upper PMS and the MS, although there should be some such stars
at least in LH~60, LH~63 and LH~72. A possible reason for this lack is the over 
subtraction of field stars on the TO. These points suggest that the differential
Hess diagrams are subject to the normalization of the stellar numbers of the field, 
which nevertheless does not represent the real background field in the observed 
regions, as well as to the simulation process for the reproduction of this real field.
Therefore, we  rely on the Hess diagrams only for demonstrating the existence of  
a true PMS population of each region, and not for quantifying its members.}

Considering the stellar distributions in the red-faint part of the CMDs of 
LH~60, LH~63 and LH~72, while the appearance of a sequence of stars in this part of the CMDs is quite 
obvious (see Figs~\ref{f:vi-cmd}, \ref{f:hessor} and \ref{f:difhess}), one may argue that this 
part of the CMD is populated as a result of the effect of differential 
reddening on MS stars, rather than by actual PMS stars. However, if we consider the 
visual extinction measurements performed in \S~\ref{s:red} and the maximum 
of the derived values given in Table~\ref{t:av}, as well as the corresponding uncertainties,  
we cannot fully assign the observed red-faint stellar sequences to reddening of the MS alone. 
In particular, while the brightest part of the red-faint stellar sequence, at $m_{\rm 814} \simeq 21$ and  $m_{\rm 555}  - 
m_{\rm 814} \simeq 1$, may be partly explained by extinction, the colors covered by the 
faint part of the sequence are completely independent from a hypothetical 
differential reddening of the MS, as well as from the measured photometric errors at these 
magnitudes. This can be directly verified on the CMDs shown in Fig.~\ref{f:vi-cmd} by considering 
 the directions of the reddening vectors.

The analysis thus far shows that the PMS feature in the CMDs and Hess diagrams 
of the observed clusters appears to be quite distinct from a well-defined MS in the cases 
of LH~60, LH~63 and LH~72. On the other hand, the case of LH~88 seems to be quite different, because   
1) we do not observe the same sequence of stars in the red faint part of the observed CMD and the 
corresponding Hess diagram of this region, and 2) the MS of LH~88 as seen in these 
diagrams is quite broader than those of the other regions and the field. 
This behavior can be explained by the higher extinction and its differential nature 
in the vicinity of LH~88. Indeed, there is a dense, thin {\sl residual}  sequence of stars to the red part of 
the LMS in the differential Hess diagram of LH~88 (Fig.~\ref{f:difhess}), which does match the loose 
broad sequence of PMS stars seen in the other regions. This sequence could be misinterpreted as an 
`old' PMS population, because the extinction measured for LH~88 (Table~\ref{t:av}) can easily span its
width. This stellar feature, thus, may
include both reddened LMS stars and PMS stars.  While we can not exclude the presence of PMS stars in 
LH\,88, their loci should be well contaminated by reddened MS stars, and therefore their identification 
is not as straightforward as in the other regions.

%\clearpage 
%%%%%%%%%%%%%%%%%%%%%%%%%%%% FIGURE %%%%%%%%%%%%%
\begin{figure*}[t!]
\centerline{\includegraphics[clip=true,width=1.0\textwidth]{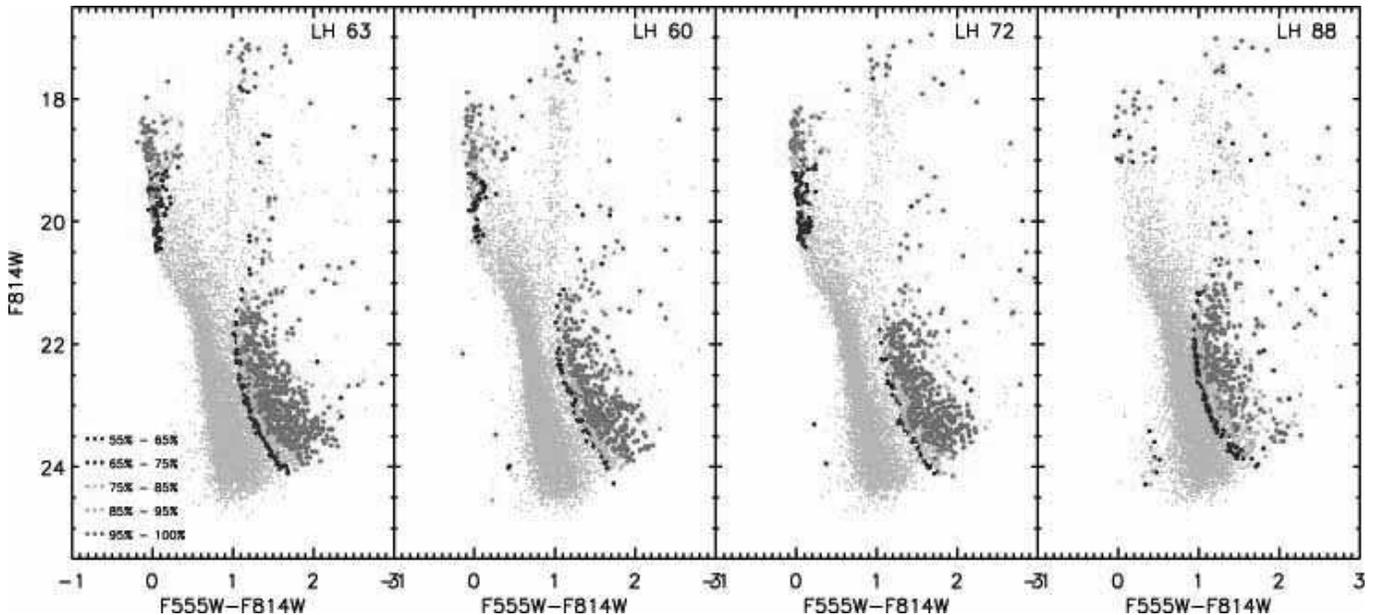}}
\caption{Field-subtracted $VI$-equivalent CMDs for the observed star-forming regions. The remaining stellar 
members are color-coded according to their membership probabilities derived from the repeated 
application of our Monte Carlo method for the field decontamination (see \S~\ref{s:mc}).
[A color version of this figure will be available in the online journal.]
\label{f:fs-cmd}}
\end{figure*}
%%%%%%%%%%%%%%%%%%%%%%%%%%%%%%%%%%%%%%%%%%%%%%

\subsection{Statistical Identification of PMS stars: Subtraction of the field stars contamination}\label{s:mc}

The stellar populations observed in the regions of the clusters are subject to 
strong contamination by the general LMC field. This contamination can be 
assessed {\sl statistically} with the use of CMDs of the LMC field (Figs.~\ref{f:fcmds}, 
\ref{f:cmdsiso} {\sl left}) and subtracted from the cluster CMDs with the application 
of a {\sl Monte Carlo} method. For this method we consider
a set of elliptical subregions around each star in the CMD of each cluster  
and the same set of subregions in the CMD of the LMC field. We then  
statistically subtract from the CMD of each system the corresponding number 
of randomly selected stars in the field CMD for each of these subregions. We 
perform 100 iterations of the Monte Carlo subtraction for each cluster, and we 
assign, thus, a percentage of membership probability, $p$, to every remaining 
star-member. The resultant field-subtracted CMDs of the clusters are shown
in Fig.~\ref{f:fs-cmd}, overlaid on the originally observed CMDs (in grey). 
The stars remaining after the subtraction process are color-coded according
to their assigned membership probability with $p \geq 55$\%. In this figure 
it is shown that, as was the case in the Hess diagrams of the previous section,  
stars located in both the faint red and bright blue part of the CMDs are flagged as 
true members of the clusters. As it is observed also in the Hess diagrams of Fig.~\ref{f:difhess}, after the field-subtraction 
there are few red {\sl evolved} stars, which are falsely flagged as cluster members. These
are residuals of the subtraction technique, which was not possible to be removed due to
natural inconsistencies between the CMDs of the regions and that of the control-field.

The large majority of residual {\sl cluster-member} stars with membership probability 
$p \geq 95$\%, occupy the PMS part of the CMDs in Fig.~\ref{f:fs-cmd}. Naturally, this 
behavior characterizes the cases of LH~60, LH~63 and LH~72, where the PMS part of the CMD 
is fully populated by cluster-member stars down to the faintest detected magnitudes. On the other 
hand, in the case of LH~88,  residual red stars with high membership probabilities correspond to somewhat brighter 
magnitudes and form a sequence connected to the residual field stars of the 
sub-giant branch and RC. Considering the high extinction  of this region, the appearance of 
these red sources in a thin sequence almost {\sl attached} to the red 
part of the LMS can be best explained by the effect of differential reddening on MS stars, rather 
than by these stars being PMS stars. The case for LH~63, which is also affected by - nonetheless lower - 
extinction, is not similar to that of LH~88, because the stars with high-membership
probability populate completely the PMS part of the CMD as in the cases of LH~60 and LH~72.  
It is interesting to note that in the cases of these two clusters, which show the
lowest extinction, there is a clear {\sl gap} between the LMS and the PMS stars, making the 
distinction of the latter more straightforward. 

Nevertheless, it should be noted that the results of the Monte Carlo subtraction
technique are subject to its parameters. In particular the resulted `clean' 
CMDs and the remaining stellar numbers depend on the size of the individual 
elliptical regions around each star considered for the random elimination of 
the field stars. Also this technique makes use of a {\sl general} control-field rather
than the {\sl actual} field at the area of each observed region. Therefore, we
treat the results of the Monte Carlo technique qualitatively and not quantitatively,
as we did for the Hess diagrams. A quantitative analysis for the determination of
the {\sl most probable} PMS member-stars of each region is performed in the next 
section. Taking into 
account the results of this section, and in 
order to eliminate the appearance of reddened field stars in our PMS samples, 
in our subsequent analysis we concentrate on the fainter part of the CMDs with 
$m_{\rm 814} \gsim 21$.

\section{Membership determination of PMS stars in the observed regions} \label{s:pmsmmb}

In this section we perform a statistical determination of the membership of the 
PMS stars in each observed star-forming region by constructing the stellar 
distributions along cross-sections of the observed CMDs. Since, as discussed above, 
at magnitudes brighter than $m_{814} \sim 21$ there is a confusion between PMS
and evolved stars, we limit our treatment to the fainter part of each CMD. We consider 
again only the stars found with photometric uncertainties $\bar{\delta_{2}} \leq 0.1$.

%\clearpage 
%%%%%%%%%%%%%%%%%%%%%%%%%%%% FIGURE %%%%%%%%%%%%%
\begin{figure}[t!]
\centerline{\includegraphics[clip=true,width=\columnwidth]{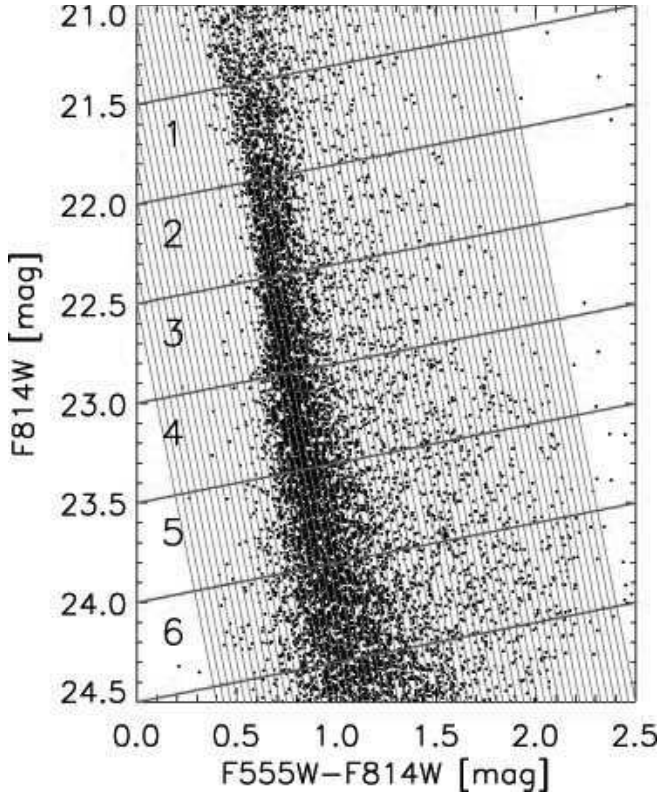}}
%\vspace*{1.truecm}
%\centerline{\large Figure is omitted due to size limitations.} 
%\centerline{\large Copy available upon request to the first author.}
%\vspace*{1.truecm}
\caption{Example of the stellar counting process for the construction of the distribution
of stars along cross-sections through the observed $VI$-equivalent CMD of LH~60.
The six selected cross-sections perpendicular to the MS and the PMS locus are indicated 
by the red lines, covering the whole faint part of the observed magnitude range. The strips 
selected for the binning of the stars along the cross-sections, and the construction of the 
corresponding stellar distributions are indicated by the blue lines, almost parallel to the MS.
[A color version of this figure will be available in the online journal.]  \label{f:cmdcuts}}
\end{figure}
%%%%%%%%%%%%%%%%%%%%%%%%%%%%%%%%%%%%%%%%%%%%%%

%\clearpage 
\subsection{Stellar distributions 
along cross-sections of the CMD}\label{s:crssct}

A more quantitative analysis of the observed PMS populations is achieved by counting 
the number of stars as a function of their ($m_{555}-m_{814}$) colors along a series of 
CMD cross-sections perpendicular to the PMS locus.  
This  method is previously established for the identification of PMS stars  
in $VI$ CMDs of Galactic star-forming regions \citep[see, e.g.,][]{sherry04}.
An example of the counting process for the construction of the stellar 
distributions along six cross-sections, selected to cover the magnitude 
range of interest, i.e., 21.5~\lsim~$m_{814}$~\lsim 24.5, is shown
in Fig.~\ref{f:cmdcuts} for LH~60. The cross-sections are indicated  by the red lines 
with their numbers marked. Stars are binned in strips perpendicular to the cross-sections
and thus almost parallel to the MS. These strips, which are also almost parallel 
to typical PMS isochrones derived from PMS evolutionary models \citep[e.g.,][]{palla99, 
siess00}, are indicated in Fig.~\ref{f:cmdcuts} by the blue lines. 

%%%%%%%%%%%%%%%%%%%%%%%%%%%% FIGURE %%%%%%%%%%%%%
\begin{figure*}[t!]
\centerline{\includegraphics[clip=true,width=1.000\textwidth]{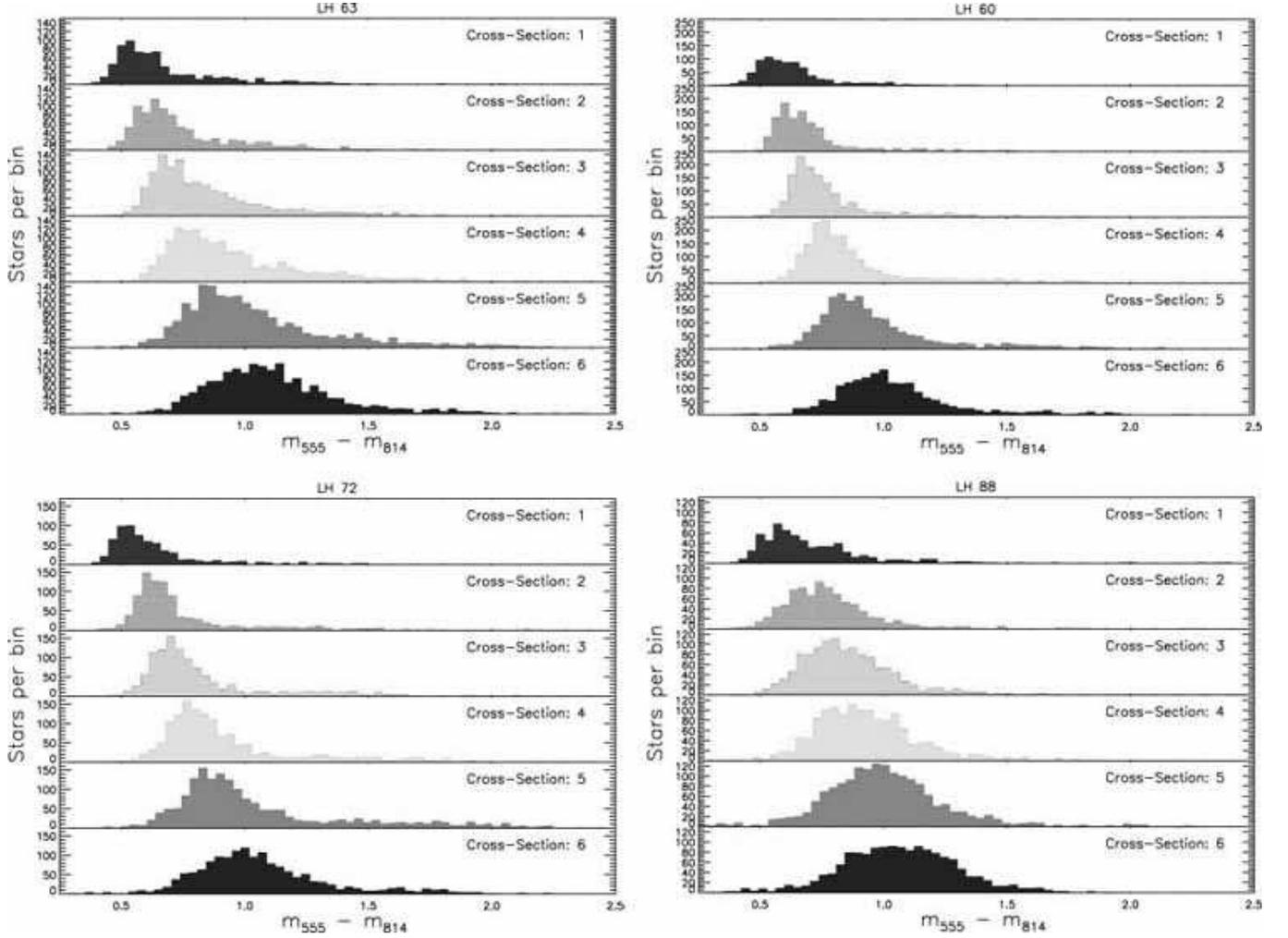}}
\caption{Stellar distributions along selected cross-sections of the CMDs of the observed regions. 
Since we are interested in the faint PMS populations of the clusters we constrain 
our analysis to six cross-sections through the CMD, covering the magnitude range 
21.5~\lsim~$m_{814}$~\lsim 24.5. Stars are counted in bins 0.035~mag wide and the 
derived distributions are plotted with respect to the corresponding average color 
($m_{\rm 555} - m_{\rm 814}$) of each bin. Different colors are used for the 
distributions of different cross-sections. Stellar numbers are corrected for incompleteness.
These plots reveal the PMS populations of the 
observed clusters as a secondary distribution red-ward from that peaked on the field 
MS stars. [A color version of this figure will be available in the online journal.] \label{f:starprof}}
\end{figure*}
%%%%%%%%%%%%%%%%%%%%%%%%%%%%%%%%%%%%%%%%%%%%%%

%\clearpage 
%%%%%%%%%%%%%%%%%%%%%%%%%%%% FIGURE %%%%%%%%%%%%%
\begin{figure}[t!]
\centerline{\includegraphics[clip=true,width=\columnwidth]{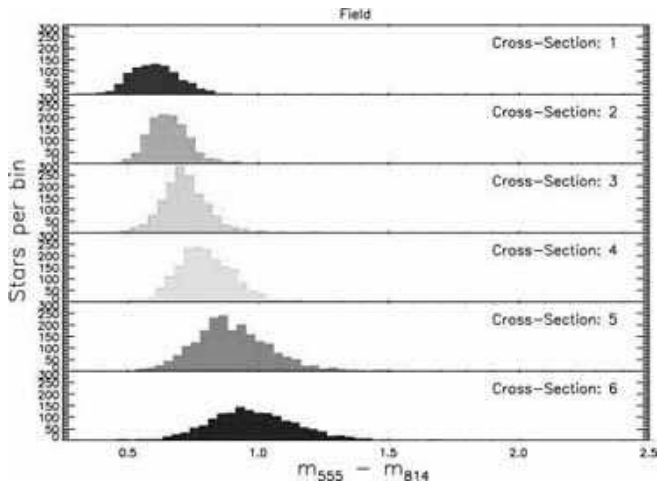}}
\caption{Same as Fig.~\ref{f:starprof} for the total population of the general LMC field. 
[A color version of this figure will be available in the online journal.] \label{f:fstarprof}}
\end{figure}
%%%%%%%%%%%%%%%%%%%%%%%%%%%%%%%%%%%%%%%%%%%%%%

The constructed stellar distributions are shown in Fig.~\ref{f:starprof}. Stellar numbers
are corrected for incompleteness according to our completeness measurements 
described in \S~\ref{s:phterr-cmp}. { Since the distribution of stars and nebula is not
uniform across each observed field, confusion may also vary across each region, being 
higher at the most crowded regions. As a consequence, photometric completeness not 
only is a function of the observed magnitudes but also depends on the position of each 
star across the field. Therefore, we corrected the numbers of stars per bin in every strip 
on a single-star basis according to the magnitude and position of each star in the 
observed field.} In  Fig.~\ref{f:starprof} it can be seen that the 
distribution of stars for each cross-section is peaked on the field MS, which comprises
the large majority of the observed stars in each target. Our aim is to identify the stellar 
distributions that correspond to the redder PMS population of the star-forming regions.
Indeed, as it can be seen for the regions LH~60, LH~63 and LH~72, their stellar distributions 
are not symmetrically peaked on the MS and are clearly extended to the red, especially
for cross-sections with $m_{814}$~\lsim 24.0.
It should be noted that, 
%according to our measurements of visual extinction in these three regions, 
the red-ward asymmetry of their stellar distributions 
{\sl cannot} be attributed to reddening alone based on our visual extinction estimates.

For comparison to these distributions we show in Fig.~\ref{f:fstarprof} the corresponding stellar distributions 
derived from the CMD of the control fields (Figs.~\ref{f:fcmds}, \ref{f:cmdsiso}). From 
the distributions of Fig.~\ref{f:fstarprof} it is shown that field stars, 
in contrast to those of the star-forming regions, are distributed {\sl symmetrically} around the 
MS. Again, the case of LH~88 fits more this behavior, rather than that of the other three 
regions of our sample. The stellar distributions in LH~88, shown in Fig.~\ref{f:starprof}, appear 
symmetrically peaked
on the MS, as in the case of the LMC field, but much broader than the general field,
most probably due to the high extinction that characterizes this region. Under these
circumstances, even if there are PMS stars in LH~88, we cannot accurately assess
their numbers. In all of the discussed distributions, the wider spread of stars for cross-sections 
corresponding to fainter magnitudes can be naturally attributed to the larger photometric 
uncertainties.

%\clearpage 
%%%%%%%%%%%%%%%%%%%%%%%%%%%% FIGURE %%%%%%%%%%%%%
\begin{figure*}[t!]
\centerline{\includegraphics[clip=true,width=0.825\textwidth]{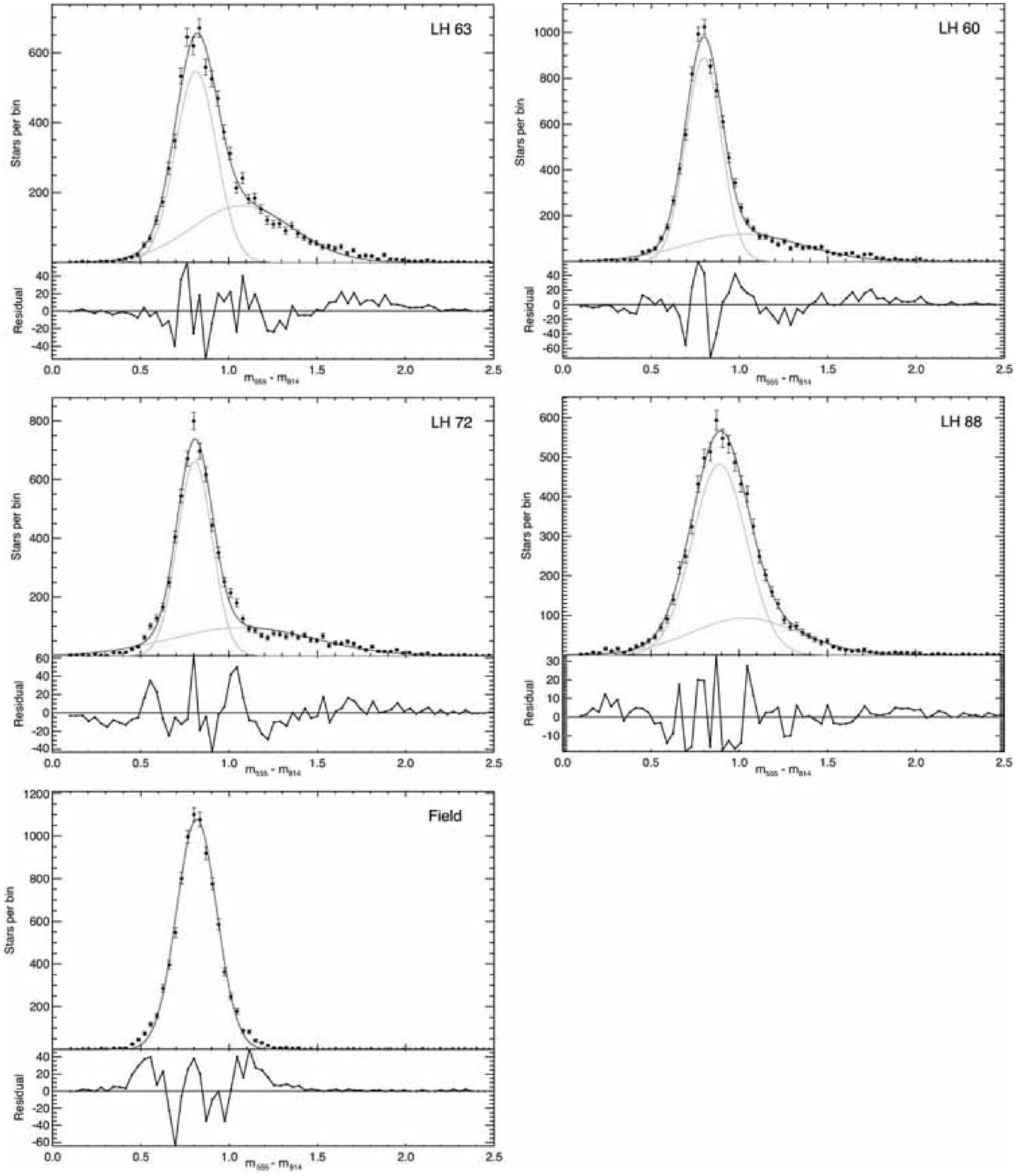}}
\caption{Cumulative stellar distributions along {\sl all} considered 
cross-sections of the CMDs for the magnitude range 21.5~\lsim~$m_{814}$~\lsim 
24.5. Stars are counted in bins 0.035~mag wide and their numbers are 
corrected for incompleteness. The distributions are plotted with respect 
to the average color ($m_{\rm 555} - m_{\rm 814}$) of each strip (bin). The 
best-fit double Gaussian function is overlaid with a red line. Each component
of this function is also drawn with orange lines. Errors represent Poisson statistics. 
Residuals are plotted below each fit, demonstrating the goodness-of-fit.
[A color version of this figure will be available in the online journal.] \label{f:starproffits}}
\end{figure*}
%%%%%%%%%%%%%%%%%%%%%%%%%%%%%%%%%%%%%%%%%%%%%%

Considering the above, the distribution of stars along cross-sections through 
the CMDs of the star-forming regions can be well represented by the sum of 
two distributions; one for the field MS stars and one of the native 
PMS stars of the regions.  For demonstration we apply such a fit to the {\sl cumulative} stellar
distribution for each region and the field. These distributions are constructed by counting the stars
in the selected strips parallel to the MS for the whole considered magnitude 
range, i.e., in all six selected cross-sections, as they are shown in Fig.~\ref{f:cmdcuts}.
In this manner, the peak of the distribution will keep following the original
peaks on the MS of the individual cross-sections. On the other hand, since the 
counting bins (strips) are not vertical to the color axis, each bin in the cumulative 
distribution does not correspond to a single color value. Therefore, we assign to each
bin the {\sl average} color index, which roughly corresponds to that of the average 
magnitude of $m_{814} \simeq 23$.

We fit the number of stars per bin of the cumulative distributions to a double Gaussian 
function of the form \begin{equation} \label{eq:gaussfit} y =  \displaystyle 
\sum\limits_{i=1}^{2} N_{i} e^{-0.5((x-\mu_{i})/\sigma_{i})^2}. \end{equation}
The first term describes the distribution of field MS stars and the
second that of the PMS stars.  We use a least-squares multiple Gaussian fit 
performed by the interactive IDL routine {\sc xgaussfit} (by Don Lindler)
to solve for six parameters, $N_{i}$, $\mu_{i}$, and $\sigma_{i}$, where $i=1$ is for
the field and $i=2$ for the PMS stars.  The constructed cumulative stellar distributions across the CMDs of the 
observed regions and the LMC field are shown in Fig.~\ref{f:starproffits}. 
The best-fitted Gaussian distributions derived from our fitting process are also
drawn; the total fit is indicated with a red line, while each of the two Gaussian 
components is drawn with an orange line. { The corresponding residuals of 
the fits are accordingly drawn in order to show the goodness of each fit.}

From the plots of this figure for LH~60, LH~63 and LH~72 one can see 
that the red component of the stellar distributions along the cross-sections of 
the CMDs is indeed important and successfully represented by a Gaussian 
distribution. In the case of LH~88 the contribution of the red component 
to the stellar distribution turns out -- as expected --  not to be significant, 
as can be seen by the fits. The cumulative distribution of stars in the field 
CMD is shown also in Fig.~\ref{f:starproffits} for comparison. Naturally, only 
a single-Gaussian fit  could be applied to this distribution. The coefficients of the 
fits  shown in Fig.~\ref{f:starproffits} are given in Table~\ref{t:gauss}. The values 
of this table show the remarkable coincidence of the primary 
peak in the distributions of the star-forming regions with the Gaussian function 
representing the general field. This coincidence, which is more prominent in 
LH~60, LH~63 and LH~72, provides additional evidence of the field origin 
of the MS population observed in the CMDs of these star-forming regions.

%\clearpage
%%%%%%%%%%%%%%%%%%%%%%%%%%%%%%%%%%%%%%%%%%%%%%%%%%%%%%%%%%%%
\begin{deluxetable}{lrrcrrc}
\label{t:gauss}
%\tablecolumns{7}
\tablewidth{0pc}
\tablecaption{Coefficients of the best-fitting double Gaussian function to the 
cumulative distributions of faint stars across the CMDs. \label{t:gauss} }
\tablehead{
\colhead{} & 
\multicolumn{3}{c}{Gaussian component 1}&
\multicolumn{3}{c}{Gaussian component 2} \\
\colhead{Region} & 
\colhead{$N_{1}$} & 
\colhead{$\mu_{1}$} &
\colhead{FWHM$_1$} & 
\colhead{$N_{2}$} & 
\colhead{$\mu_{2}$} &
\colhead{FWHM$_2$} 
} 
\startdata
LH 63 &  547 & 0.82  & 0.27  & 162 & 1.08 & 0.68\\
LH 60 &  888 & 0.80  & 0.24  & 118 & 1.03 & 0.78\\
LH 72 &  663 & 0.81  & 0.23  &  96 & 1.08 & 0.94\\
LH 88 &  483 & 0.89  & 0.36  &  93 & 1.03 & 0.71\\
Field   & 1071 &  0.82 &  0.27&   & & 
\enddata
%\tablenotetext{*}{}
%\tablecomments{The given $\pm$ values are the measured spreads in $A_V$, not the uncertainties in the measured values.}
\end{deluxetable}
%%%%%%%%%%%%%%%%%%%%%%%%%%%%%%%%%%%%%%%%%%%%%%%%%%%%%%%%%%%

This fitting process is performed for the stellar distributions along each of the 
selected cross-sections for each observed CMD. The distribution of the field stars 
per se is not interesting, but it is crucial because  the number of PMS stars depends 
upon the extrapolation of the field star distribution through the PMS locus.  
The total number of PMS stars along a cross-section is determined by 
the normalization of the Gaussian fit to the PMS distribution, 
$N_{2,j}$ (where $j=1,...,6$ is the cross-section number), 
and by $\sigma_{2,j}$, which  defines the width of the Gaussian.  
The full width at half maximum ({FWHM}) of the
PMS population is 2${\sigma_{2,j}}{\sqrt{2ln(2)}}$.  The $\mu_{2,j}$ parameter
is the color $m_{555}-m_{814}$ of the peak of the PMS distribution along a
cross-section.  Since each cross-section is a predefined line across
the CMD, $\mu_{2,j}$ specifies the magnitude that corresponds to 
the peak of the PMS distribution. { The coefficients of the best-fit 
second Gaussian component, representative of the PMS population, for
each cross-section are given in Table~\ref{t:gauss2}.}

We fit the field and PMS distributions jointly, and we derive the total number of 
PMS stars from  the second Gaussian component of each fit  by
assigning a membership probability to each star.  
This probability is determined by 
the ratio of the second Gaussian component of the fit to the 
total fit  along each cross-section through the CMD, i.e., the 
sum of both Gaussian components of the fit. Membership 
probabilities calculated in this manner vary from 0\% to the blue 
of the PMS locus to $>$90\% near the peak of the PMS distribution 
and for redder colors in all cross-sections.  We thus derive 1~815 stars 
in LH~63, 1~230 in LH~60, 1~223 in LH~72 and only 314 in LH~88
with probabilities of cluster membership $p \geq 95\%$. These numbers are 
completeness corrected.

%\clearpage 
%%%%%%%%%%%%%%%%%%%%%%%%%%%% FIGURE %%%%%%%%%%%%%
\begin{figure}[t!]
\centerline{\includegraphics[clip=true,width=\columnwidth]{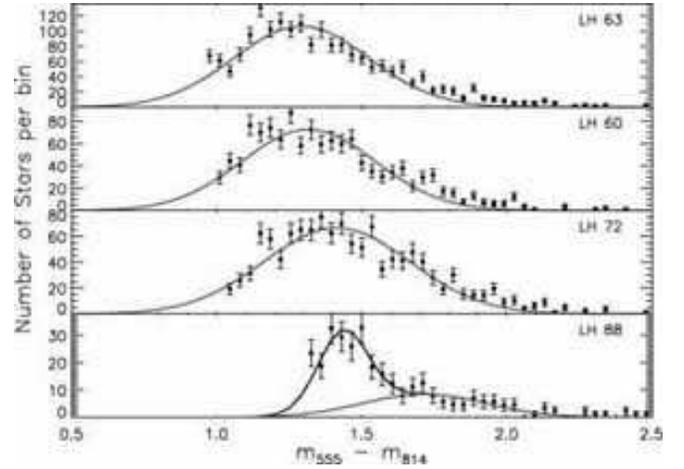}}
\caption{Cumulative stellar distributions across the CMDs of the observed 
regions for the stars found with membership probability $p \geq 95\%$. 
The PMS populations of LH~60, LH~63 and LH~72 demonstrate an extraordinary 
similarity, which possibly indicates their common characteristics. On the 
other hand the distribution for LH~88 demonstrates the difficulty in disentangling 
the {\sl true} stellar population of the region, since stars with high membership 
probability seem to represent reddened MS stars and not PMS stars, which 
are possibly represented by the second (faint) peak. 
[A color version of this figure will be available in the online journal.] \label{f:pmsprof}}
\end{figure}
%%%%%%%%%%%%%%%%%%%%%%%%%%%%%%%%%%%%%%%%%%%%%%

%\clearpage
%%%%%%%%%%%%%%%%%%%%%%%%%%%%%%%%%%%%%%%%%%%%%%%%%%%%%%%%%%%%
\begin{deluxetable*}{crrcrrcrrcrrc}
\tabletypesize{\scriptsize}
\label{t:gauss2}
%\tablecolumns{7}
\tablewidth{0pc}
\tablecaption{Coefficients of the best-fitting Gaussian component 2 of the fits to each of the 
individual cross-sections distributions of faint stars across the CMDs. \label{t:gauss2} }
\tablehead{
\colhead{} & 
\multicolumn{3}{c}{LH~63}&
\multicolumn{3}{c}{LH~60}&
\multicolumn{3}{c}{LH~72}&
\multicolumn{3}{c}{LH~88} \\
\colhead{Cross-section} & 
\colhead{$N_{2}$} & 
\colhead{$\mu_{2}$} &
\colhead{FWHM$_2$} & 
\colhead{$N_{2}$} & 
\colhead{$\mu_{2}$} &
\colhead{FWHM$_2$} &
\colhead{$N_{2}$} & 
\colhead{$\mu_{2}$} &
\colhead{FWHM$_2$} &
\colhead{$N_{2}$} & 
\colhead{$\mu_{2}$} &
\colhead{FWHM$_2$} 
} 
\startdata
1 &   17&    0.79&    0.49&    11&    0.85&    0.45&      9&    1.00&    0.58&       8&   1.17&   0.43\\
2 &   21&    0.93&    0.51&    14&    0.90&    0.45&    14&    0.92&    0.66&    14&    0.93&   0.45\\
3 &   53&    0.86&    0.41&    31&    0.89&    0.42&    14&    1.09&    0.76&    24&    0.98&   0.49\\
4 &   37&    1.09&    0.60&    24&    1.07&    0.70&    18&    1.17&    0.86&    18&    1.14&   0.53\\
5 &   31&    1.25&    0.76&    32&    1.13&    0.73&    20&    1.32&    1.05&    20&    1.09&   0.81\\
6 &   14&    1.45&    0.73&    16&    1.30&    0.89&    13&    1.46&    0.74&       6&    1.61&   0.14
\enddata
%\tablenotetext{*}{}
%\tablecomments{The given $\pm$ values are the measured spreads in $A_V$, not the uncertainties in the measured values.}
\end{deluxetable*}
%%%%%%%%%%%%%%%%%%%%%%%%%%%%%%%%%%%%%%%%%%%%%%%%%%%%%%%%%%%

This particular statistical technique for the 
determination of cluster membership probabilities of PMS stars 
is quite self-consistent, because it makes use of the LMC field as it is 
observed {\sl within} each region and not of a generic remote 
field as in the case of the {\sl Monte Carlo} field subtraction presented 
in \S~\ref{s:mc}. Moreover, the use of the cross-sections distributions
for the determination of the most probable PMS stars is independent of 
the reddening measurements in each region. It is interesting to 
note that this technique returned two times more PMS member 
candidates (with $p \geq 95\%$) than the {\sl Monte Carlo} field 
subtraction, except in the 
case of LH~88, where the derived numbers are comparable. The 
reason for this difference lies on the elimination by the 
{\sl Monte Carlo} technique of the red wing of the LMS from the CMDs of the 
star-forming regions as field candidates. Since this technique is based
on the iterative random subtraction of candidate field contaminants within 
individual CMD-regions around each observed star, the subtraction of 
different stars in each repetition results in low membership probabilities 
for {\sl all} stars in CMD-regions where field stars are expected to be located. 
On the other hand, the cross-sections technique, being 
dependent solely on the actual CMD positions of the observed 
stars and their distributions, returns a {\sl purely probabilistic} membership
determination for each star, independent of the appearance of field stars 
or not in its surrounding CMD-region.

\section{Discussion} \label{s:disc}

In Fig.~\ref{f:pmsprof} we show the cumulative distributions along the CMD 
only for the PMS stars of each region, i.e., for those stars found with a 
membership probability $p \geq$~ 95\%. These distributions are constructed 
as those for all observed stars by counting them in strips (bins) parallel 
to the MS, as described in \S~\ref{s:crssct}. As for those distributions, 
the color index per bin is the average, corresponding to that at magnitude 
$m_{814} \simeq 23$. The distributions of Fig.~\ref{f:pmsprof} show an 
extraordinary similarity for the cases of LH~60, LH~63 and LH~72. Indeed, 
the fitting coefficients of the best-fitted Gaussian functions (drawn in red) 
are almost identical. Specifically, the peaks $\mu$ of the fitting functions 
are equal to 1.32, 1.29 and 1.41, and the corresponding widths $\sigma$ are 
0.23, 0.23 and 0.25 for LH~60, LH~63 and LH~72 respectively. The case of 
LH~88 is again an outlier and it is shown only for completeness.
The CMD distribution of the `best candidate' stellar members for LH~88 is not
represented by a single but by two Gaussians, with the first component of 
the fit (drawn in blue) being very narrow. This clearly suggests that the 
stars flagged in the previous section as best candidates for members of the 
star-forming region, and especially those at bluer colors, are in fact 
misidentified reddened MS stars. The smaller red Gaussian component 
may represent traces of the PMS population of the system, which is contaminated 
by reddened evolved stars, but this result cannot be conclusive. 

The CMD distributions of the stellar members of LH~60, LH~63 and LH~72  
show that these PMS stars cover a broad area of the CMDs of these regions. Indeed, 
the loci of PMS stars in the CMDs of Milky Way star-forming regions often show 
a widening, which could be evidence for age-spread \citep[][]{briceno07}.  
This is also observed in young clusters and associations of both  
the Magellanic Clouds imaged with \emph{HST} \citep[see e.g., ][]{gouliermis-eslab07}, 
and our results so far show that the regions investigated here are not exceptions.  
However, simulations performed previously by us have shown that characteristics
of PMS stars,  such as variability, binarity and circumstellar extinction can 
cause considerable scatter of the positions of the PMS stars in the CMD, 
giving false evidence of an age-spread within the systems \citep{hennekemper08, 
dario10}. Consequently we cannot attribute the observed CMD broadening of PMS 
stars solely to age-spread. 

{ Nevertheless,  the distributions of Fig.~\ref{f:pmsprof} convey valuable information
about the most probable ages of the systems and their indicative age-spreads. Taking the
peaks of the cumulative distributions at `face value' one can conclude that LH~63 should 
be somewhat older than LH~60 and LH~72, because it corresponds to a bluer color, and
thus to an older age. However, considering the PMS stellar distributions in the individual 
cross-sections across the CMDs for all three regions, they are all consistent with each other, 
peaking at colors along isochrones of ages between  $\sim$~3 and 5~Myr,
with the cases of LH~60 and LH~72 showing to be somewhat closer to  the
$\sim$~3~Myr isochrone. As far as the widths of the PMS distributions are concerned, they 
overlap with each other so well so that one cannot conclude any specific age difference  among the 
three clusters. The derived age span of $\sim$~3 to 5~Myr is in excellent agreement 
with the age of the star-forming region LH~95, located at the north-western 
edge of the super-giant shell LMC~4, derived by a self-consistent  
age determination technique developed by us \citep{dario10}. In a subsequent study 
we will apply this technique to the data presented here, in order to  constrain 
further the ages of the systems. It is worth noting that   taking  also the widths of
the distributions at `face value', they correspond to a significant spread between 2 and 10 Myr.
This result, however, is to be tested in our forthcoming study.}

Another factor to be considered as responsible for the CMD broadening of PMS stars 
is differential reddening. However, while the PMS stars are expected to be somewhat 
dislocated due to reddening, our measurements of visual extinction, discussed in 
\S~\ref{s:red}, show that differential reddening cannot either be fully responsible 
for the scattering of PMS stars in the CMDs of these three regions. Our measurements 
show that $A_V$ in all investigated systems, except of LH~88, is quite low with a small 
variation with a maximum of $A_V \simeq 0.68$~mag (for LH~63), corresponding roughly to 
color excess of $E({\rm F555W} - {\rm F814W}) \simeq 0.27$~mag, much smaller than the 
FWHM of the observed PMS broadening. On the other hand,  LH~88 suffers from the highest 
extinction, with a maximum of $A_V \simeq 1.5$~mag, which appears to be adequate to account 
for the observed CMD spread of stars.

In conclusion, for the regions LH~60, LH~63 and LH~72, the cross-sections 
distributions of the PMS stars with $m_{\rm 814}$~ \gsim~21.5 seem 
to be quite similar to each other in respect to their total numbers of stars, 
to their peaks and widths, as well as to the average color index at which the 
peak in stellar numbers appears. These extraordinary similarities clearly imply 
that the PMS stars found in three different young clusters, 
embedded in star-forming regions along the periphery of LMC~4, have possibly similar characteristics, and
probably share a common star formation history.

\section{Summary and Concluding Remarks}\label{s:concl}

In this paper we present the first part of our investigation of PMS stellar 
populations in the LMC. Our targets of interest are four star-forming regions,
LH~60, LH~63, LH~72 and LH~88, located along the rim of the super-giant shell LMC~4. 
We present the reduction of the multi-wavelength images taken with 
{\sl HST}~WFPC2 within our program GO-11547 (PI: D. Gouliermis), and our 
photometric analysis. We determine the accuracy and completeness of our
photometry, and we measure the visual extinction towards the regions. We 
identify the stellar populations comprised in the observed  fields in 
terms of the constructed CMDs. We evaluate the contribution by the field 
stellar populations with the use of archival WFPC2 images of the local 
LMC field. We make use of our rich photometric catalogs in the F555W and 
F814W ($V$- and $I$-equivalent) filters to assess the stellar content of 
the clusters embedded in the observed regions.

We demonstrate that the true stellar content of the star-forming regions 
comprises both bright young stars observed at the UMS of their CMDs and 
faint stars still in the PMS phase of their evolution. An outlier from this 
general behavior is the region of LH~88, the high extinction of which does not 
allow a clear identification of its true populations. The identification
of the PMS stars in the observed clusters is performed through the assessment of 
the contamination of the observed populations by the stellar content of the LMC field.
This is achieved qualitatively by the construction of the differential Hess diagrams of the 
observed regions (\S~\ref{s:hessd}), and the derivation of the ``clean'' 
CMDs of the clusters after the statistical subtraction of the contaminant field population 
from the observed CMDs (\S~\ref{s:mc}).

The probabilistic identification of the PMS stars in the observed regions is further 
performed, in a quantitative manner, by the distributions of the stars along 
cross-sections of the observed CMDs (\S~\ref{s:pmsmmb}). We particularly focus 
on the faint PMS populations of the clusters, and we constrain our analysis to the 
fainter part of the CMDs with $m_{\rm 814} \gsim 21.5$, which is less contaminated 
by evolved field populations. We construct the number distributions of 
the stars across each observed CMD along six selected magnitude ranges and we fit 
these distributions with a two Gaussian components. The first component 
represents the MS field stars observed in each region, while the second corresponds 
to the PMS member stars of the region. For each star we assign a membership probability, 
$p$,  derived from the ratio of the second fitted Gaussian to the sum of the two 
components along each cross-section. We isolate the stars with $p \geq$~95\%, as the best 
candidates of being members of the star-forming regions. All these candidates are PMS 
stars. The CMD distributions of these stars show an extraordinary similarity to one
another for the regions LH~60, LH~63 and LH~72, suggesting similar characteristics.
{ Considering that the peaks of these distributions represent the most probable ages
of the regions, their similarity also suggests that all three regions may share 
a common recent star formation history. This result is quite important, since  
these regions are located at different parts of the boundaries of the super-giant 
shell LMC~4, and therefore our findings suggest that 
star formation around the shell may have occurred almost simultaneously. 
Nevertheless,} these distributions show a definite widening of the loci of faint PMS stars in the 
observed $VI$-equivalent CMDs, { which should be further investigated}.

The spread of PMS stars along the color axis in the CMDs of star-forming regions in the 
Magellanic Clouds is a well-documented phenomenon, which demonstrates that the particular 
location of a PMS star in the CMD is a very complex function of its intrinsic properties
 \citep[see, e.g.,][]{gouliermis-eslab07}. Low-mass PMS stars in such regions, being the 
counterpart of Galactic T~Tauri stars, suffer from rotational variability, non-periodic 
variability due to accretion, circumstellar extinction, and binarity. These characteristics 
dislocate the stars from their original positions in the CMD in an unpredictable manner, 
which may be misinterpreted as an age-spread within the host stellar system. The only 
accurate approach to quantify this effect and to confirm or refute the spread in ages is 
through  detailed simulations of synthetic CMDs and their comparison to the 
observed ones. We are currently undertaking such an investigation with the accurate statistical 
determination of the masses and ages of the identified PMS stars  
through two self-consistent techniques established by us \citep[see][]{dario09,
dario10}. Our subsequent investigation of the observed star-forming regions will include 
the multi-wavelength characterization of the bright stellar content of the clusters
\citep[see, e.g.,][]{romaniello02}, the determination of circumstellar accretion of 
bright PMS stars through their $U$- and H{\alp}-excess emission 
\citep[e.g.,][]{guido2010}, and the identification of embedded massive candidate 
{\sl young stellar objects} and {\sl ultra-compact {\sc Hii} regions} via the synergy of our 
observations with those in longer wavelengths \citep[e.g.,][]{gruendl09}.

\acknowledgements
We are indebted to 
the unknown referee for her/his comments, which helped us improve this 
investigation significantly. D.A.G. kindly acknowledges 
the German Aerospace Center (DLR) and the German Federal Ministry for
Economics and Technology (BMWi) for their  support through grant 50~OR~0908. 
Support for this work was provided by NASA through grant number HST-GO-11547
from the SPACE TELESCOPE SCIENCE INSTITUTE, which is operated by the
Association of Universities for Research in Astronomy, Inc., under NASA
contract NAS5-26555.

%A.E.D, M.R., Y.-H.C, R.A.G and N.P. acknowledge the support of NASA through grant HST-GO-11547. 

%Based on observations
%made with the NASA/ESA {\em Hubble Space Telescope}, obtained from the
%data archive at the Space Telescope Science Institute. STScI is operated
%by the Association of Universities for Research in Astronomy, Inc. under
%NASA contract NAS 5-26555. 

Facilities: HST

%German Research Foundation (Deu\-tsche For\-schungs\-ge\-mein\-schaft, DFG) through 
%grant GO~1659/1-2

\end{document}